\documentclass[a4paper,twocolumn,11pt,accepted=2025-04-23]{quantumarticle}
\pdfoutput=1
\usepackage[utf8]{inputenc}
\usepackage[english]{babel}
\usepackage[T1]{fontenc}
\usepackage{amsmath}
\usepackage{subfigure}
\usepackage{psfrag,graphicx}
\usepackage{pict2e}
\usepackage{dcolumn}
\usepackage{amsmath,amssymb,braket}
\usepackage{bm}
\usepackage{color}
\usepackage{latexsym}
\usepackage{epstopdf}
\usepackage{color}
\usepackage[english]{babel}
\usepackage{soul}
\usepackage{comment}
\UseRawInputEncoding
\usepackage{amsfonts}
\usepackage{bm}
\usepackage{natbib}
\usepackage{bbold}

\usepackage{enumerate}
\definecolor{myblue}{rgb}{0.2,0.2,0.8}
\definecolor{myzard}{cmyk}{0,0,0.05,0}
\definecolor{mywhite}{rgb}{1,1,1}
\definecolor{mywhite}{rgb}{1,1,1}
\definecolor{myred}{rgb}{1,0.,0.3}
\usepackage[colorlinks=true,citecolor=myblue,linkcolor=myred]{hyperref}

\definecolor{darkgreen}{rgb}{0.0, 0.4, 0.26}

\definecolor{mygrey}{gray}{0.35}
\definecolor{myblue}{rgb}{0.2,0.2,0.8}
\definecolor{myzard}{cmyk}{0,0,0.05,0}
\definecolor{mywhite}{rgb}{1,1,1}
\definecolor{mywhite}{rgb}{1,1,1}
\definecolor{myred}{rgb}{1,0.,0.3}

\def\be{\begin{equation}}
\def\ee{\end{equation}}
\def\ba{\begin{align}}
\def\enda{\end{align}}
\def\bi{\begin{itemize}}
\def\ei{\end{itemize}}

\def\beq{\begin{equation}}
\def\beq{\begin{equation}}
\def\eeq{\end{equation}}

\newcommand{\mean}[1]{\langle #1\rangle}

\usepackage{tikz}
\usepackage{lipsum}

\begin{document}

\title{Topological, multi-mode amplification induced by non-reciprocal, long-range dissipative couplings}

\author{Carlos Vega}
\affiliation{Institute of Fundamental Physics IFF-CSIC, Calle Serrano 113b, 28006 Madrid, Spain.}
\orcid{0000-0002-2633-9585}
\author{Alberto Mu\~{n}oz de las Heras}
\affiliation{Institute of Fundamental Physics IFF-CSIC, Calle Serrano 113b, 28006 Madrid, Spain.}
\orcid{0000-0003-3836-6107}
\author{Diego Porras}
\affiliation{Institute of Fundamental Physics IFF-CSIC, Calle Serrano 113b, 28006 Madrid, Spain.}
\orcid{0000-0003-2995-0299}
\author{Alejandro Gonz\'{a}lez-Tudela}
\affiliation{Institute of Fundamental Physics IFF-CSIC, Calle Serrano 113b, 28006 Madrid, Spain.}
\orcid{0000-0003-2307-6967}

\maketitle

\begin{abstract}
  Non-reciprocal couplings or drivings are known to induce steady-state, directional, amplification in driven-dissipative bosonic lattices. This amplification phenomenon has been recently linked to the existence of a non-zero topological invariant defined with the system's dynamical matrix, and thus, it depends critically on the couplings' structure. In this work, we demonstrate the emergence of unconventional, non-reciprocal, long-range dissipative couplings induced by the interaction of the bosonic chain with a chiral, multi-mode channel, and then study their impact on topological amplification phenomena. We show that these couplings can lead to topological invariant values greater than one which induce topological, multi-mode amplification and metastability behaviour. Besides, we also show how these couplings can also display topological amplifying phases that are dynamically stable in the presence of local parametric drivings. Finally, we conclude by showing how such phenomena can be naturally obtained in two-dimensional topological insulators hosting multiple edge modes.
\end{abstract}

\section{Introduction}
\label{sec:introduction}

Non-reciprocity, the property of physical processes of depending on the direction in which they occur, is both a topic of fundamental and applied interest. On the fundamental side, it has been studied since the origin of quantum optics in the context of cascaded quantum systems~\cite{gardiner1993driving, carmichael1993quantum}, characterized by a unidirectional coupling between different nodes. It is also at the heart of recently studied puzzling phenomena such as the non-Hermitian skin effect~\cite{alvarez2018non,yao2018edge,lin2023topological}, dynamical phase transitions in classical active matter~\cite{fruchart2021non,avni2023}, anomalous quantum optical responses~\cite{gong2022a,gong2022b}, or unconventional photon-mediated interactions~\cite{gong2022a,gong2022b,rocatti2024, roccati2022exotic, du2023giant}, among others. On a more applied perspective, non-reciprocity can be used to improve quantum sensing protocols~\cite{Lau2018,McDonald2020,budich2020,parto2023,sarkar2023} or to induce chiral light-matter interactions~\cite{lodahl17a} with applications in the dissipative preparation of many-body entangled states~\cite{stannigel2012driven,ramos14a,pichler15a,ramos16a}, the generation of complex states of light~\cite{pichler17a,mahmoodian2020,ostfeldt2022,mittal18a,Harari2018,kartashov19,Bandres2018,longhi2018,Secli2019,Amelio2020,zeng2020}, or the routing~\cite{metelmann2018nonreciprocal,jalas2013and, macdonald2018a,Karakaya2020,wonlnik2020,deBernardis2023} and processing of quantum information~\cite{Schrinski2021}. For these reasons, strong efforts are being directed towards engineering non-reciprocal couplings with different methods~\cite{metelmann2015nonreciprocal,ozawa19a,wang2024dispersive, verhagen2017optomechanical,wang2023quantum,Soro2022,Sanchez-Burillo2020c,guimond2020}, which have already resulted in experimental realizations in various platforms~\cite{Roushan2017,Ruesink2016,wanjura2023,delpino2022,kannan2023,slim2024,Owens2022, reisenbauer2024non}.

One of the most attractive applications of such non-reciprocal couplings is the development of directional amplifiers~\cite{deak2012,caloz2018, pucher2022atomic, mercier2019realization}. Such systems are capable of amplifying an input signal in one direction while avoiding backscattering in the other, with numerous applications in very disparate fields ranging from radioastronomy~\cite{smith2013low} to microwave quantum technologies~\cite{blais2021circuit}. 

Along this direction, arrays of parametric amplifiers ~\cite{macklin2015twpa} are one of the ways of achieving directional amplification, with a gain that potentially increases with system size. 
However, even though current devices show some degree of directionality, backscattering into the detected object is still significant thus posing limitations in certain applications like measurement of qubits in quantum computers~\cite{Esposito2021}.

Motivated by this challenge, several works~\cite{peano16a,Porras2019,Wanjura2020,Wanjura2021,Flynn2020,Flynn2921a,
Ramos2021, gomez2022bridging, GomezLeon2022b,GomezLeon2023,Ramos2022a, wanjura2023quadrature,Tian2023} 
have recently put forward a new concept based on a topological dissipative phase that leads to low-noise, full directional amplification with 
a \emph{built-in} protection against imperfections. 
This protection is rooted in the emergence of a quantized topological invariant defined with the dynamical matrix of the driven-dissipative-photonic lattice~\cite{Porras2019,Wanjura2020}, which is directly connected to the existence of exponentially-amplified steady-states. For this reason, such phenomenon has been dubbed as \emph{topological amplification}. The proposals to obtain such robust amplification have been mostly based on non-reciprocal couplings of the Hatano-Nelson~\cite{Hatano1996} type, such as in Refs.~\cite{Porras2019,Wanjura2020,gomez2022bridging,GomezLeon2022b,Tian2023,Ramos2021}, or via parametric cross-Kerr couplings~\cite{Ramos2022a,GomezLeon2023}.

In this work, we demonstrate the existence of a novel topological amplification regime featuring multiple amplification channels. The origin of this regime lies in a different type of non-reciprocal couplings that we also demonstrate to appear when the photonic lattice modes interact via a chiral, multi-mode waveguide~\cite{Vega2023TopologicalQED,Skirlo2014,Skirlo2015}. Beyond this channel multiplicity, we also show that these non-reciprocal tunnelings give rise to both a dynamical metastability of these driven-dissipative phases and a stabilization of such topological amplification using only local parametric drives. The manuscript is structured as follows:  In Sec.~\ref{sec:effective_interactions}, we derive the non-reciprocal couplings mediated by chiral, multi-mode waveguides, and emphasize their differences with the ones appearing in the standard chiral quantum optical regime~\cite{lodahl17a}. In Sec.~\ref{sec:driven_dissipative_phases}, we study the driven-dissipative topological phases that these non-reciprocal tunnelings can induce with both incoherent and parametric pumping. This includes Sec.~\ref{sec:dynamics}, where we study the transient behaviour before arriving to these steady-state phases to illustrate the emergence of metastability. Finally, in Sec.~\ref{sec:implementation}, we also discuss how such couplings and phenomena can be experimentally probed by harnessing the chiral, edge modes of two-dimensional topological insulators, e.g., built from microwave resonators~\cite{Owens2022}. 

\section{Non-reciprocal, multi-mode couplings}
\label{sec:effective_interactions}

\begin{figure}[t!]
    \includegraphics[width=0.99\columnwidth]{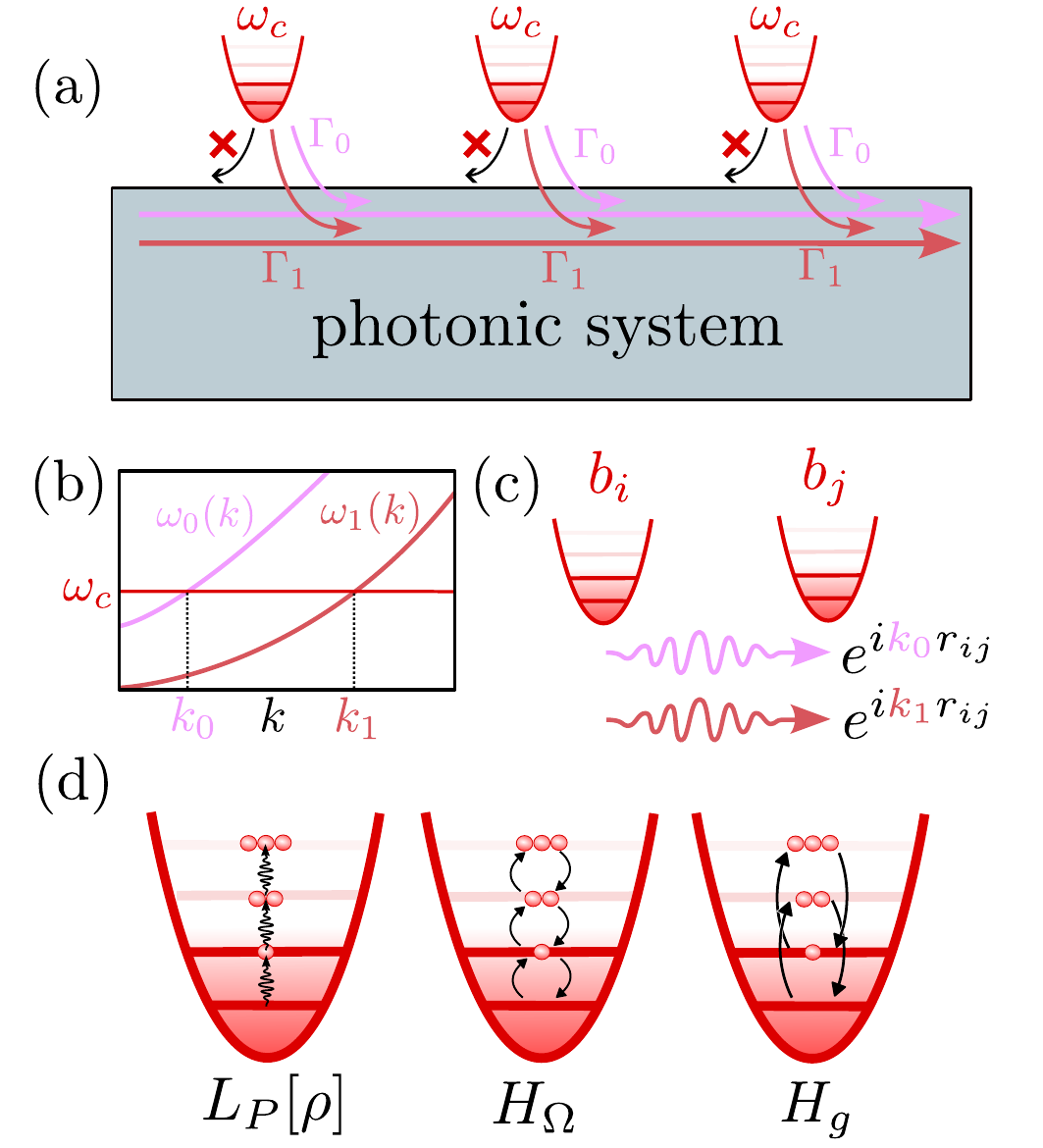}
    \caption{\textbf{Scheme of the system.} (a) The photonic lattice is assumed to be composed by a cavity array with equal frequencies $\omega_c$ and described by bosonic harmonic oscillator operators, $b_{i}^{(\dagger)}$, depicted in red. We also assume that the cavities only exchange excitations via a chiral, multi-mode waveguide channel. We also assume that cavity modes are coupled to each waveguide mode with rate $\Gamma_\ell$, exemplified in a two-mode waveguide as $\Gamma_0$ and $\Gamma_1$ in the figure. (b) Schematic representation of the band-structure of the multi-mode waveguide for the two-mode scenario with energy dispersions $\omega_\ell(k)$. The energy of the cavities $\omega_c$ defines the resonant momenta of each of the channels $k_\ell$ that dominate the transfer of excitations. (c) Adiabatically eliminating the photon field results in effective chiral tunnelings in which each of the modes features a different propagating phase, i.e., $e^{ik_\ell r}$. Thus, compared to the standard chiral quantum optical regime~\cite{lodahl17a}, here the position between the cavities cannot be gauged away. (d) Schematic representation of the excitation mechanisms used in this work: the local incoherent pumping $L_P[\rho]$ in Eq.~\eqref{eq:pumping_dissipator}, the coherent drive in Eq.~\eqref{eq:coherent_drive}, and the local parametric driving in Eq.~\eqref{eq:para}.}
    \label{fig:scheme}
\end{figure}

Throughout this work, we consider a coupled cavity array description for the photonic lattice as depicted in Fig.~\ref{fig:scheme}(a). The lattice is composed by $N$ local cavity modes, which we assume to have the same frequency $\omega_c$, described by bosonic operators $b_i^{(\dagger)}$. Their free Hamiltonian is then given by $H_S=\omega_c\sum_{i}b_i^\dagger b_i$, where we set $\hbar=1$ as we will do for the rest of the paper.

The key difference of this work with respect to the others in the literature is that we assume that the non-reciprocal couplings between the different cavities do not occur directly, but rather via the interaction of the cavities with a multi-mode waveguide as depicted in Fig.~\ref{fig:scheme}(a). The Hamiltonian of multi-mode waveguide modes can be written as:
\begin{equation}
H_B = \sum_\ell \int dk\;\omega_\ell(k)A_{\ell,k}^\dagger A_{\ell,k}  \;,\label{eq:HB}
\end{equation}
where $\ell$ is the integer index that denotes the different waveguide modes, which runs from $\ell= 1, 2,\dots, n_\mathrm{modes}$, $\omega_\ell(k)$ is the energy dispersion associated to the $\ell$-th mode, and $A_{\ell,k}^{(\dagger)}$ are the bosonic operators describing the waveguide modes with momenta $k$. On the other hand, the cavity-waveguide coupling can be written in general as:
\begin{equation}
H_\text{int} = \sum_i\sum_\ell\sum_{k} g_k(r_i)\; b_i^\dagger A_{\ell,k} + \text{H.c.}\;,\label{eq:Hint}
\end{equation}
with $r_i$ being the position where the $i$-th cavity mode couples to the waveguide, and $g_k(r_i)$ its coupling strength to the waveguide photons with momentum $k$, described by a bosonic operator $A_{\ell,k}^{(\dagger)}$ which obeys the commutation relations $[A_{\ell,k}, A_{\ell',k'}^\dagger]=\delta_{\ell,\ell'}\delta(k-k')$.

The non-reciprocity of the couplings can appear either because the waveguide modes are chiral themselves, that is, there exist only left or right moving modes as it happens at the edges of two-dimensional topological insulators hosting multiple, chiral edge modes~\cite{Vega2023TopologicalQED,Skirlo2014,Skirlo2015}; or because, while both left and right moving modes exist, the cavities only couple to one of them, e.g., using optical spin-orbit coupling~\cite{lodahl17a,Bliokh2015}. Mathematically, the first situation implies that the $k$-integral in Eq.~\eqref{eq:HB} only runs for $k\lessgtr 0$, whereas the second scenario implies that $g_k(r_i)$ in Eq.~\eqref{eq:Hint} is different from zero only for $k\lessgtr 0$. Irrespective of the mechanism, adiabatically eliminating the waveguide modes under the Born-Markov conditions, see Appendix~\ref{sec:master_equation} for a complete derivation, one arrives at an effective master equation for the $b_i$-cavity modes which reads
\begin{align}
\frac{\partial \rho}{\partial t}=i\left[\rho (H_S+H_\text{eff}^\dagger) -(H_S+H_\text{eff})\rho\right]+\mathcal{J}[\rho]\,\label{eq:meq}
\end{align}
with $\mathcal{J}[\rho]$ encapsulating the quantum jump terms, and $H_\text{eff}$ the effective non-Hermitian evolution where one explicitly sees the emergence of both coherent ($J_{ij}$) and incoherent ($\Gamma_{ij}$) couplings between the cavity modes,
\begin{equation}
H_\text{eff}=\sum_{ij} \left(J_{ij}-i\frac{\Gamma_{ij}}{2}\right)b_i^\dagger b_{j}\;,
\label{eq:effective_Hamiltonian}
\end{equation}
with:
\begin{widetext}
\begin{equation}
J_{ij}-i\frac{\Gamma_{i,j}}{2}
=-i\sum_\ell \frac{\Gamma_\ell}{2}e^{ik_\ell(r_i-r_j)-|r_i-r_j|/l_\kappa}\left(1+\text{sign}(r_i-r_j)\right)\;.    
\end{equation}
\label{eq:chiral_multimode_self_energy}
\end{widetext}

Here, we assume the sum in $\ell$ is performed over $n_\text{modes}$ waveguide modes, $\Gamma_\ell$ is the effective decay rate of the cavity mode to the $\ell$-th mode, $k_\ell$ is the resonant momenta of each of the channels defined by $\omega_\ell(k_\ell)=\omega_c$, and $l_\kappa$ is a parameter that we introduce to account for the finite propagation length of the waveguide modes due to either absorption losses or other imperfections. In this work, we assume that $l_\kappa$ has a finite value, which guarantees spectral convergence in the thermodynamic limit~\cite{defenu2023long}. Such waveguide-mediated tunnelings have several combined features that make them qualitatively different to any other platform that has been studied in the literature in this context:
\begin{itemize}
    \item First, as expected due to the chirality of the modes, the couplings are non-reciprocal due to the $\text{sign}(r_i-r_j)$. In particular, if $r_i<r_j$ they are strictly zero, that is, the excitations only tunnel through positions $r_i>r_j$.
    \item They allow to hop beyond nearest-neighbouring sites. In particular, if $l_k\rightarrow \infty$ they are infinite-ranged, as it is typical in other waveguide QED systems~\cite{Sheremet2023WaveguideCorrelations}.
    \item Critically, if $n_\text{modes}\neq 1$, the cavity positions, $r_i$, cannot be gauged away due to the different propagating phases of the waveguide modes in the waveguide channels, i.e., $e^{i k_\ell r}$, see Fig.~\ref{fig:scheme}(c). Note that this is very different from the standard chiral quantum optical regime~\cite{lodahl17a} where only the ordering, and not the position between the cavities, matters. We recover the standard chiral quantum optical regime in Eq.~\eqref{eq:chiral_multimode_self_energy} for $n_\mathrm{modes}=1$ or if $k_\ell\equiv k_0$ for all $\ell$.
\end{itemize}

\begin{figure}[t!]
    \includegraphics[width=0.99\columnwidth]{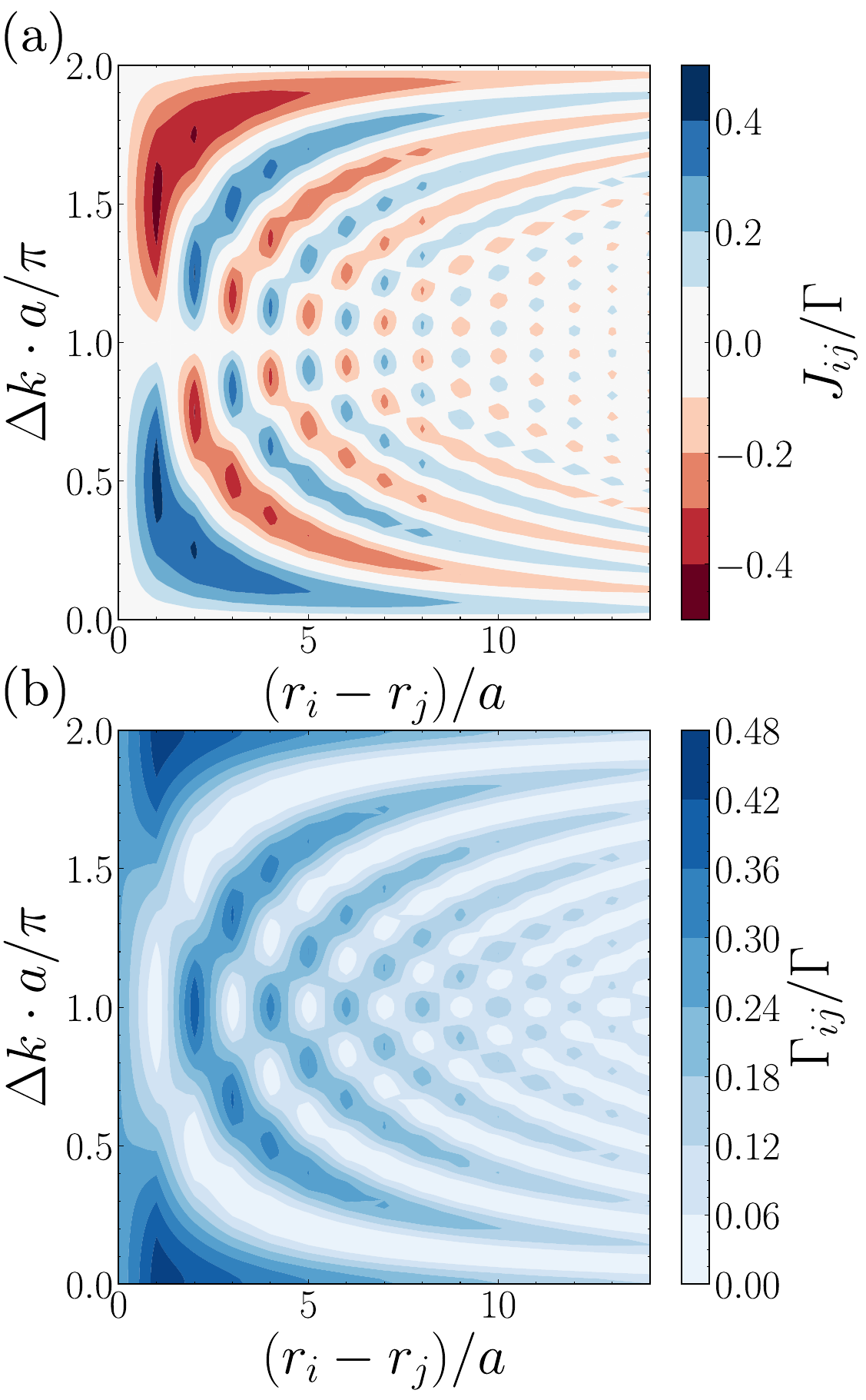}
    \caption{\textbf{Non-reciprocal couplings}. (a) Effective coherent $J_{ij}$ and (b) incoherent couplings $\Gamma_{ij}$ from Eq.~\eqref{eq:chiral_multimode_self_energy} for a two-mode waveguide situation with $\Gamma_0=\Gamma_1=\Gamma/2$ and a propagation length $l_\kappa/a=10$. Both quantities are plotted as a function of the distance between mode $i$ and mode $j$ (only if $r_i\geq r_j$, otherwise $J_{ij}=\Gamma_{ij}=0$), and the effective multi-mode phase difference $\Delta k\cdot a$.}
    \label{fig:Role_of_Delta_k}
\end{figure}

In Fig.~\ref{fig:Role_of_Delta_k} we plot an example of such long-range, non-reciprocal couplings, $J_{ij}$ (top) and $\Gamma_{ij}$ (bottom) in a two-mode scenario for an equally-spaced set of cavities $r_j\equiv j\cdot a$, with $j\in\mathbb{N}$, and a finite propagation length $l_k/a=10$. We consider the dependence on both the distance between the cavities $r_i-r_j$, restricted to $r_{i}\geq r_j$, otherwise, we have $J_{ij}=\Gamma_{ij}=0$) and the effective multi-mode phase difference of the two channels $\Delta k\cdot a$ with $\Delta k=k_0-k_1$. There, we observe more explicitly how the emitters position cannot be gauged away unless $\Delta k\cdot a=0,2\pi$, resulting in different oscillatory phases between the coherent and incoherent part of the couplings. Such oscillations allow one to find regimes in which the hoppings between longer distances are larger than nearest-neighbouring ones, which has strong implications for the topological amplification phenomena as we will see in the next section. A very clear example of that occurs for the $\Delta k\cdot a=\pi$ situation where the photons from the two channels arrive in anti-phase into the neighbouring sites resulting in a cancellation of both $J_{ij}$ and $\Gamma_{ij}$ at the neighbouring sites. In fact, $J_{ij}\equiv 0$, for any $i$ and $j$ in that regime, resulting into a bipartite, pure, dissipative coupling.

\section{Driven-dissipative topological phases and amplification}
\label{sec:driven_dissipative_phases}

In the previous section, we demonstrate the appearance of the qualitatively different non-reciprocal couplings written in Eq.~\eqref{eq:chiral_multimode_self_energy}. In this section, we study the consequences of such couplings in the context of topological amplification~\cite{peano16a,Porras2019,Wanjura2020,Wanjura2021,Flynn2020,Flynn2921a,Ramos2021,GomezLeon2022a,GomezLeon2022b,GomezLeon2023,Ramos2022a,Tian2023}. For that, we need to include some type of driving or pumping into the system such that the steady-state in the system is not the trivial one, that is, the vacuum state of the cavities. In this work, we consider three possible mechanisms of adding excitations into the system schematically depicted in Fig.~\ref{fig:scheme}(d):
\begin{itemize}
    \item Local incoherent pumping, with rate $P$, described by an additional term into the master equation Eq.~\eqref{eq:meq}:
    \begin{align}
    L_P[\rho]=\frac{P}{2}\sum_i \left(2b_i^\dagger \rho b_i-b_i b_i^\dagger\rho-\rho b_i b_i^\dagger\right)\,.
    \label{eq:pumping_dissipator}
    \end{align}

    \item Local coherent driving with amplitude $\Omega$ and frequency $\omega_L$ described a Hamiltonian:
    \begin{align}
    H_{\Omega}=\Omega \sum_i (b_i e^{i\omega_L t}+b_i^\dagger e^{-i\omega_L t})\,.
    \label{eq:coherent_drive}
    \end{align}

    \item Local parametric drive with amplitude $g_s$ and frequency $2\omega_p$ described by a Hamiltonian:
    \begin{align}
    H_{g}=g_s \sum_i \left(b_i^2 e^{i2\omega_p t}+(b_i^\dagger)^2 e^{-i2\omega_p t}\right)\,.
    \label{eq:para}
    \end{align}
    
\end{itemize}

Thus, going to a rotating frame with the parametric driving frequency $2\omega_p$ and assuming $\omega_L=2\omega_p$, the master equation which describes the driven-dissipative photonic lattice reads:
\begin{align}
\frac{d\rho}{dt}=&-i\left(H_*\rho - \rho H_*^\dagger\right)  
+\mathcal{J}(\rho)\;,
\label{eq:master_equation}
\end{align}
with $H_*$ being the effective non-Hermtian Hamiltonian including both the effective couplings and potential drivings and pumping terms:
\begin{align}
H_* = &\Delta\sum_{i}b_i^\dagger b_i+\sum_{ij}\left(J_{ij}-i\frac{\Gamma_{ij}}{2}\right)b_i^\dagger b_j \nonumber\\
+&i\frac{P}{2}\sum_i \;b_i^\dagger b_i+\Omega_i(b_i^\dagger+b_i )+g_s(b_i^{\dagger 2}+b_i^2)\,,
\label{eq:H_nh_driving}
\end{align}
where we define $\Delta\equiv \omega_c-2\omega_p$, and $\mathcal{J}(\rho)$ being the quantum jump terms due to the pumping and incoherent couplings:

\begin{equation}
\mathcal{J}(\rho) = \sum_{ij}\Gamma_{ij}b_i\rho b_j^\dagger  +P\sum_{i}b_i^\dagger\rho b_i    \;.
\end{equation}

Recent works~\cite{peano16a,Porras2019,Wanjura2020,Wanjura2021,Flynn2020,Flynn2921a,Ramos2021,GomezLeon2022a,GomezLeon2022b,GomezLeon2023,Ramos2022a,Tian2023} have found that under some conditions, such excitation mechanisms can result into non-trivial photonic steady-states in which the coherent input signal $\Omega$ is exponentially amplified. In the next section, we will see how such amplification can be connected to a topological invariant that is defined using the dynamical matrix of the system.

\subsection{Review of topological amplification}

\subsubsection{With incoherent gain}

Let us start considering the situation with $g_s=0$ and $\Delta=0$. In that case, the evolution of the coherences of the system, i.e., $\mathbf{b}\equiv (\langle b_1\rangle,...,\langle b_N\rangle)^T$, is governed by the following set of linear differential equations:
\begin{equation}
\frac{d\mathbf{b}}{dt} = -i\mathbb{H}\mathbf{b}+i\mathbf{\Omega}\;,
\label{eq:eq_of_motion_b}
\end{equation}
with $(\mathbb{H})_{ij}=J_{ij}-i\frac{\Gamma_{ij}}{2}+i\frac{P}{2}\delta_{i,j}$. It is important to highlight that (the inverse of) $\mathbb{H}$ also governs the spatial shape of the steady-state since $d\mathbf{b}_\text{ss}/dt=0$ implies that:
\begin{align}
\mathbf{b}_\text{ss} = \mathbb{H}^{-1}\mathbf{\Omega}\,.
\label{eq:ss_solution}
\end{align}
which can be rewritten using the singular value decomposition of $\mathbb{H}=USV^\dagger$ (where $U$ and $V$ are unitary matrices and $S_{nm}=s_n\delta_{nm}$ is a diagonal matrix containing the singular values $s_n\geq 0$) as follows:
\begin{equation}
\langle b_j\rangle_\text{ss}=\sum_i\sum_n V_{jn}\frac{1}{s_n}U_{in}^\star\Omega_i  \;.
\label{eq:steady_state_coherences}
\end{equation}

As first realized by Ref.~\cite{Porras2019}, such singular value decomposition can also be used to define a topological character of the steady-states by realizing that the singular values $s_n$ are also the eigenvalues of the following Hermitian matrix $\mathcal{H}$:
\begin{equation}
\mathcal{H}=\begin{pmatrix}
0 & \mathbb{H}\\
\mathbb{H}^\dagger & 0
\end{pmatrix}\;,
\label{eq:doubled_H}
\end{equation}
dubbed as the \emph{doubled Hamiltonian}. In particular, it can be proven that:
\begin{align}
\mathrm{eig}\left[\mathcal{H}\right]=\lambda_n=\pm s_n\,,\label{eq:eig}
\end{align}
as we remind again in Appendix~\ref{sec:eigvals_and_singular_values}. The key point to make the connection with standard topological band-theory is to realize that since $\mathcal{H}$ has an artificial chiral symmetry, it typically falls into the BDI or AIII classes of the Altland-Zirnbauer tenfold topological classification~\cite{ryu2010topological}. This means that, in one-dimension, such Hamiltonian can support non-trivial topological phases characterized by an integer topological invariant and manifested by the presence of zero-energy edge states. The associated singular values $s_n\to0$ will dominate the spatial dependence of the steady-state since, as we observe in Eq.~\eqref{eq:steady_state_coherences}, their contribution to the steady-state is weighted by $s_n^{-1}$. 

Let us now define more explicitly the invariant that characterizes these phases. As it occurs in standard topological classification, such invariants are defined in the bulk, i.e., by imposing periodic boundary conditions in the photonic lattice. Within those conditions, one can diagonalize the effective part of $H_*$ using the plane-wave expansion of the operators $b_k=(1/\sqrt{2\pi})\int dr\;e^{-ikj}b_j$ as follows:
\begin{align}
\sum_{ij}\left(J_{ij}-i\frac{\Gamma_{ij}}{2}\right)b_i^\dagger b_j 
+i\frac{P}{2}\sum_i \;b_i^\dagger b_i\rightarrow \sum_k h(k)b_k^\dagger b_k\,.
\end{align}

With that $h(k)$, one can rewrite the $2\times 2$ blocks of the doubled Hamiltonian in momentum space as follows:
\begin{equation}
\mathcal{H}(k) = 
\begin{pmatrix}
0 & h(k)\\
h^\star(k) & 0
\end{pmatrix} \;.
\end{equation}

From this form, it follows that the topological invariant can be computed as (see Refs.~\cite{Porras2019,Wanjura2020} and Appendix~\ref{sec:winding_number} for a complete explanation):
\begin{equation}
W =\frac{1}{4\pi i}\int_\text{BZ}dk\;\text{Tr}\left(\tau_z\mathcal{H}(k)^{-1}\partial_k\mathcal{H}(k)\right)    \;,
\label{eq:winding_number_matrix}
\end{equation}
with $\tau_z$ being a matrix representation of the artificial chiral symmetry of the doubled Hamiltonian $\mathcal{H}$ i.e. $\tau_z\mathcal{H}\tau_z=-\mathcal{H}$. Besides, in this case with no parametric driving ($g_s=0$), the calculation of the winding number can be further simplified to:
\begin{equation}
W = \frac{1}{2\pi i}\int_\text{BZ}dk\;\partial_k\log h(k) \;,
\label{eq:winding_number_scalar}
\end{equation}
as proven in Appendix~\ref{sec:winding_number}. In this scalar form, we recover the standard interpretation of the winding number as the number of times the vector $\left(\mathrm{Re}\left[h(k)\right],\mathrm{Im}\left[h(k)\right]\right)$ winds around the origin as $k$ sweeps the Brillouin zone. 
This form of the winding number connects topological amplification to the concept of point-gap topology in non-Hermitian systems, see \cite{Kawabata2019}.
Using that formalism, topological amplifying phases with $W=1$ have been found for both nearest-neighbour~\cite{Porras2019, Wanjura2020, Ramos2021, GomezLeon2023} or infinite-range non-reciprocal hoppings~\cite{Tian2023} as well as phases with $W=2$ in short-range neighbor couplings~\cite{brunelli2023restoration, GomezLeon2023}.

\subsubsection{With parametric driving}

Let us now focus on the situation with $g_s\neq 0$ in Eq.~\eqref{eq:H_nh_driving}. In that case, one must upgrade the description in Eq.~\eqref{eq:eq_of_motion_b} because the coherence dynamics couple both $\mean{b_i}$ and $\mean{b_j^\dagger}$. This changes the dimension of $\mathbb{H}$ from being a $N\times N$ matrix to a $2N\times 2N$ one. Besides, when imposing periodic boundary conditions, the parametric driving induces an effective coupling between positive and negative momenta, see Appendix~\ref{sec:eigvals_and_singular_values} for a complete derivation, resulting into:
\begin{align}
H_{*}\rightarrow &\sum_k \left(h(k)+\Delta\right)b_{k}^\dagger b_k +  \sum_k \left(h(k)+\Delta\right)b_{-k}^\dagger b_{-k} \nonumber\\
& + g_s \sum_k \left(b_{k}^\dagger b_{-k}- b_{k}^\dagger b_{-k}\right)\,.
\end{align}

Thus, the doubled $k$-dependent Hamiltonian is defined in $4\times 4$ blocks as follows:
\begin{widetext}
$$\mathcal{H}(k) = 
\begin{pmatrix}
0 & 0 & h(k)+\Delta & g_s\\
0 & 0 & -g_s & -h^\star(k)-\Delta\\
h^\star(k)+\Delta & g_s & 0 & 0\\
-g_s & -h(k)-\Delta & 0 & 0
\end{pmatrix}\;.$$
\end{widetext}
while the definition of the topological invariant remains the same as in Eq.~\eqref{eq:winding_number_matrix}. Eq.~\eqref{eq:para} only includes local parametric couplings, something that simplifies previous schemes such as \cite{GomezLeon2023,Ramos2022a}, where non-trivial winding numbers where obtained in the presence of cross-Kerr couplings of the form $b_i b_j + b_i^\dagger b_j^\dagger$.

\subsection{Driven-dissipative topological phase diagram with non-reciprocal, long-range couplings}

\begin{figure}[t!]
    \centering
    \includegraphics[width=0.99\columnwidth]{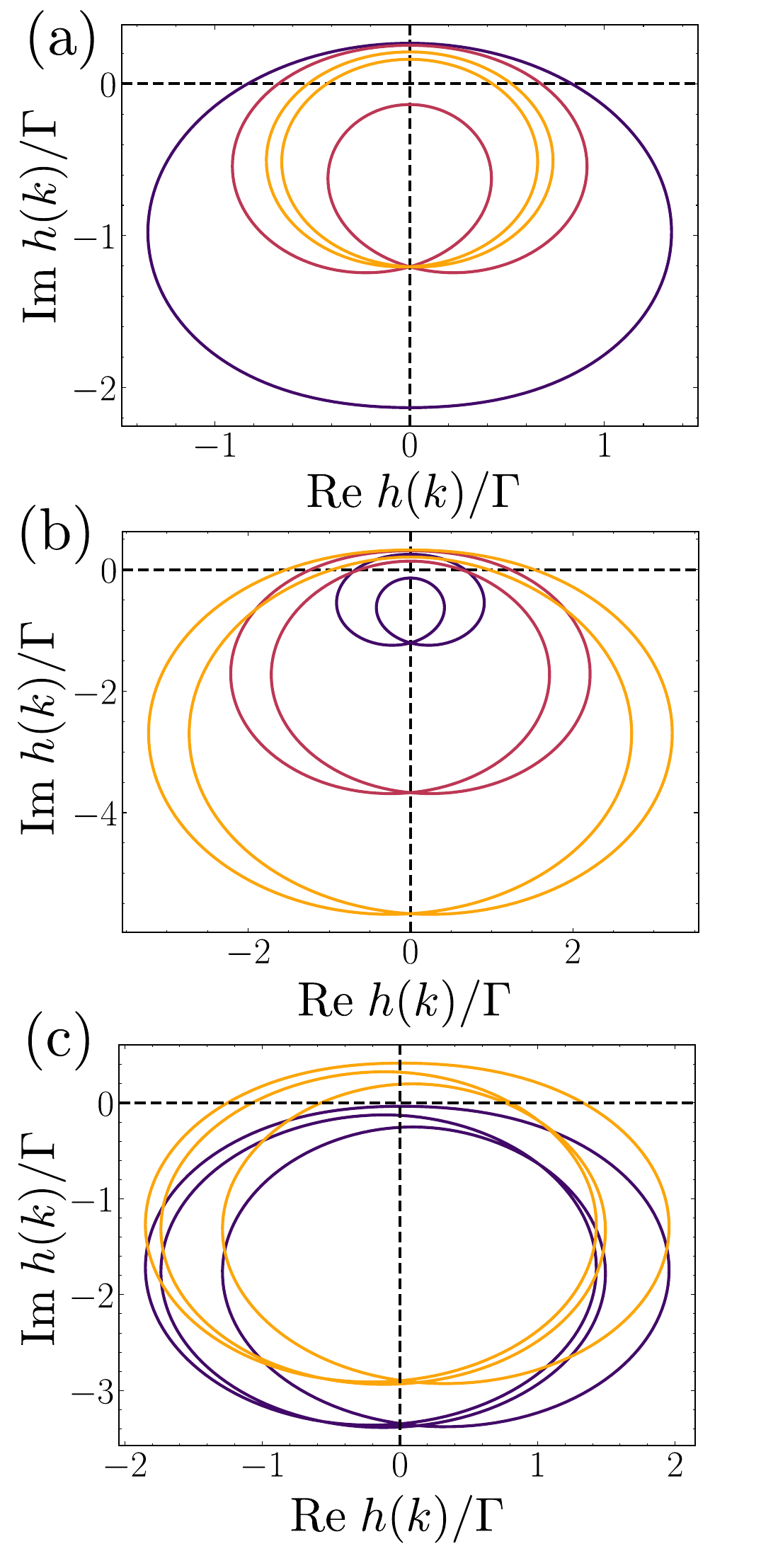}
    \caption{\textbf{Impact of model parameters on the movement of winding loops}.
    Real and imaginary part of $h(k)$ defined in Eq.~\eqref{eq:nonHermitian_spectrum} as $k$ swipes the Brillouin Zone $(-\pi,\pi)$ for several configurations. (a) Two-mode waveguide with $l_\kappa/a=3$, $P/\Gamma=0.9$. We show the cases $\Delta k\cdot a=0$ and $W=1$ (purple), $\Delta k\cdot a=\pi/10$ and $W=1$ (red), and $\Delta k \cdot a = 9\pi/10$ and $W=2$ (yellow). (b) Two-mode waveguide with $P/\Gamma=0.9$, $\Delta k\cdot a = \pi/2$. Different propagation lengths are considered: $l_\kappa/a=3$ and $W=1$ (purple), $l_\kappa/a=8$ and $W=2$ (red), and $l_\kappa/a=12$ and $W=2$ (yellow). (c) Three-mode waveguide with $(k_0a,k_1a,k_2a)=(0, \pi/2, -\pi/3)$, $l_\kappa/a=3$, for different incoherent pump rates $P/\Gamma=0$ and $W=0$ (purple) and $P/\Gamma=0.9$ and $W=3$ (yellow). In all cases, the coupling to every mode is equal i.e. $\Gamma_\ell=\Gamma/n_\text{modes}$.}
    \label{fig:winding_visualization}
\end{figure}

Let us now consider the impact of the non-reciprocal couplings derived in Section~\ref{sec:effective_interactions}. Interestingly, the form of $h(k)$ using the $J_{ij},\Gamma_{ij}$ of Eq.~\eqref{eq:chiral_multimode_self_energy} can be analytically obtained (see Appendix~\ref{sec:non_Hermitian_Hamiltonian_diagonalization}):
\begin{equation}
h(k) = i\frac{P-\Gamma}{2}-i\sum_\ell\frac{\Gamma_\ell e^{i(k_\ell-k)a-a/l_\kappa}}{1-e^{i(k_\ell-k)a-a/l_\kappa}}\;,
\label{eq:nonHermitian_spectrum}
\end{equation}

where $\Gamma\equiv \sum_\ell \Gamma_\ell$. This allows us, e.g., in the $g_s=0$ situation, to plot in Fig.~\ref{fig:winding_visualization} the movement of the vector $\left(\mathrm{Re}\left[h(k)\right],\mathrm{Im}\left[h(k)\right]\right)$ as $k$ runs over the Brillouin zone with the different key parameters of the photonic lattice:

\begin{figure*}[t!]
    \includegraphics[width=0.99\textwidth]{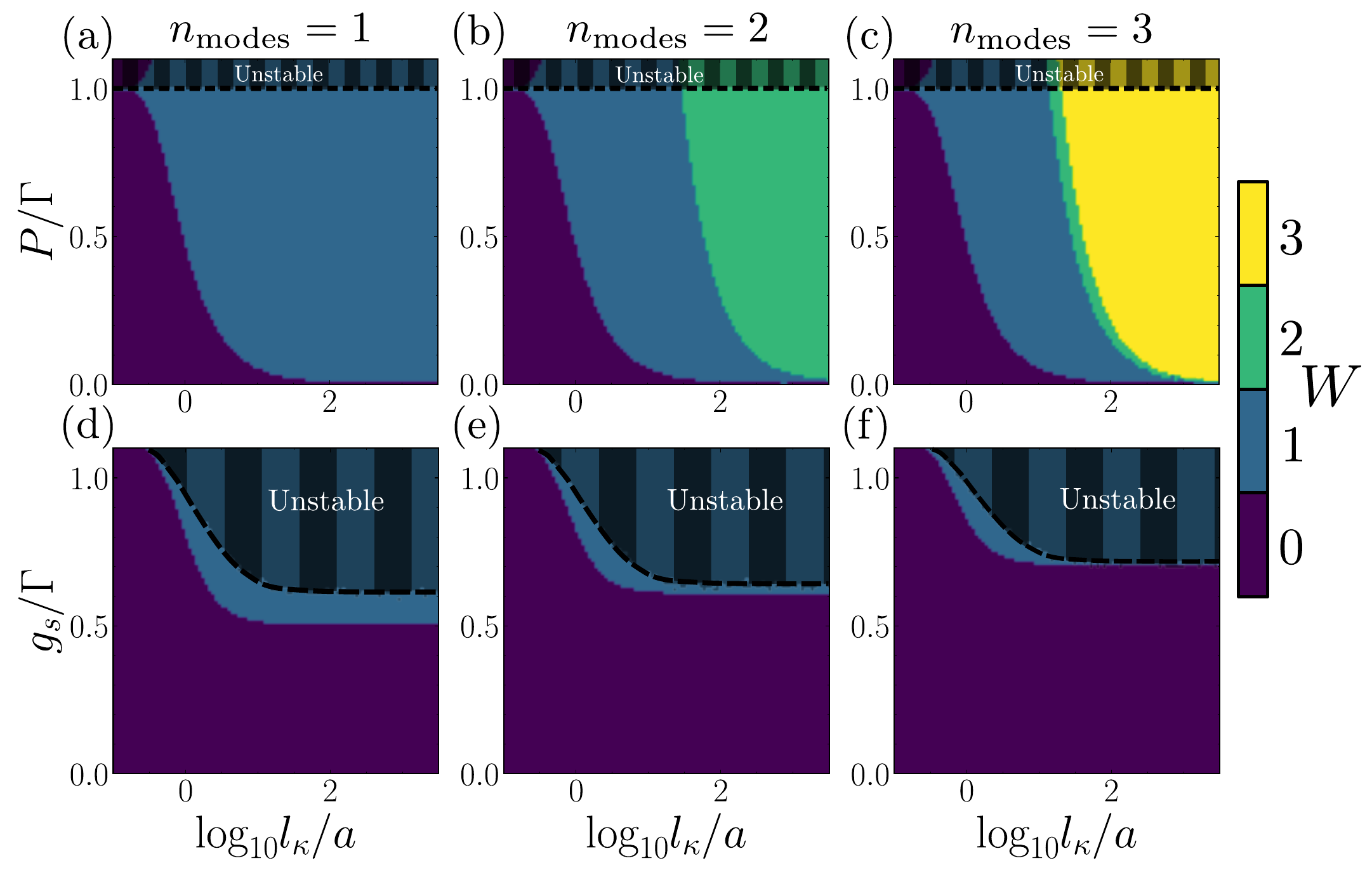}
    \caption{\textbf{Driven-dissipative topological phase diagrams}. Figures represent the winding number $W$ characterizing steady-state amplification for different propagation lengths $l_\kappa$, by adding either an incoherent pump (a, b, c) or a local parametric drive with detuning $\Delta/\Gamma=1$ (d, e, f). Shaded areas indicate dynamically unstable regions. Each column represent the phase diagram with an associated number of modes $n_\text{modes}$ of the chiral waveguide. We are considering a waveguide with (a, d) single-mode with $l_\kappa/a=$ (b, e) two-mode with $(k_0a, k_1a)=(\pi/2, \pi/3)$ (c, f) three-mode waveguide with $(k_0a, k_1a, k_2a)=(\pi/2, \pi/3, \pi/8)$. In all cases, the decay rate onto each mode is equal i.e. $\Gamma_\ell=\Gamma/n_\text{modes}$, and we define $\Gamma$ as the total decay rate $\Gamma=\sum_\ell \Gamma_\ell$}
    \label{fig:phase_diagram}
\end{figure*}

\begin{itemize}
    \item \emph{Effective multi-mode phase difference.} In Fig.~\ref{fig:winding_visualization}(a) we illustrate the winding dependence of a two-mode waveguide as a function of the multi-mode phase difference, $\Delta k \cdot a$, assuming both a fixed propagation length $l_\kappa/a=3$ and pumping rate $P/\Gamma=0.9$, being $\Gamma$ the total decay rate over the channels. There, we see how for $\Delta k \cdot a=0$ (purple line), the curve only winds once. This is expected because it recovers the single-mode scenario where it is known that only $W=1$-phases are found. For $\Delta k \cdot a=\pi/10$ (red line), the curve starts to feature a double loop, however, it is not big enough so that the second loop whirls around the origin, and thus we find still $W=1$. Finally, when $\Delta k \cdot a=9\pi/10$ (yellow line), the two loops encircle the origin, featuring then a $W=2$ phase. This emphasizes the critical role of the multi-mode nature to obtain multiple amplifying channels.

    \item \emph{Finite propagation length.} In Fig.~\ref{fig:winding_visualization}(b), we initially place ourselves in a two-mode situation with $\Delta k \cdot a=\pi/2$, $l_\kappa/a=12$ and  $P/\Gamma=0.9$ featuring a winding number $W=2$ (yellow line), and then study the impact of decreasing the propagation length. As we see in Fig.~\ref{fig:winding_visualization}, when $l_\kappa/a=8$ (red line) the loops shrink their size, however, they are still able to encircle the origin, thus keeping the $W=2$. However, when the range is short enough, i.e., $l_\kappa/a=3$ (purple line), one of the circles shrinks to the point where it does not enclose the origin any more, and the phase changes from $W=2$ to $W=1$.

    \item \emph{Incoherent pumping.} Finally, in Fig.~\ref{fig:winding_visualization}(c) we consider the effect of the incoherent pumping rate $P/\Gamma$ for a three-mode situation with $(k_0a,k_1a,k_2a)=(0, \pi/2, -\pi/3)$, $l_\kappa/a=3$. We observe that if $P=0$ (purple line) even though the three-mode waveguide induces a three-loop structure in the curve, it is not able encircle the origin, such that $W=0$. However, choosing $P/\Gamma=0.9$ (yellow line) the loops are displaced to a position where all of them enclose the origin, and thus $W=3$.
\end{itemize}

After having understood the critical role of the different terms in the movement and generation of the winding loops, in Fig.~\ref{fig:phase_diagram}(a,b,c) we plot the complete phase diagram resulting from this movement as a function of the propagation length, $l_\kappa$, in the horizontal axis, and the incoherent pumping rate $P$ for three different situations with $n_\mathrm{modes}=1,2,3$, respectively. There, we see very clearly how the number of modes sets an upper bound to the potential winding number $W$ that can be found in the system, which we formally prove in Appendix~\ref{sec:winding_number}. Remarkably, in all the situations we analyze, we predict the emergence of all the different topological phases available by choosing an appropriate combination of $l_\kappa$ and $P$. Besides, we also study the stability of the model by analyzing when the imaginary part of the eigenvalues of $\mathbb{H}$ is positive, and thus become unstable. We indicate the unstable regions with a striped shading in the phase diagram, demonstrating that one can find stable, topological phases for all the possible winding numbers.

In Figs.~\ref{fig:phase_diagram}(d-f) we do a similar analysis but now setting $P=0$ and studying the dependence in the vertical axis on the parametric driving amplitude $g_s$. Using the general expression of the winding number $W$ of Eq.~\eqref{eq:winding_number_matrix} we find that with this driving there are only non-trivial, stable phases with $W=1$ irrespective of the number of modes. Despite the absence of multi-mode amplification in this scenario, the generation of the stable topological phase $W=1$ with local parametric driving is also a consequence of the non-reciprocal, long-range hoppings $J_{ij},\Gamma_{i,j}$, since similar phases have only been predicted so far with cross-Kerr couplings~\cite{GomezLeon2023,Ramos2022a}.

\subsection{Multi-mode, topological amplification: steady-state}
\label{sec:multi_mode_amplification}

After studying formally the phase diagram through the calculation of the winding number $W$ in the \emph{bulk} and finding phases with $W>1$, now we study the physical consequences of such phases in finite systems, i.e., with open boundary-conditions. In particular, in this section we focus on the steady-state features, such as its spatial shape and momentum distribution.\\

Due to the bulk-boundary correspondence, when $W\neq 0$ we expect two consequences:
\begin{itemize}
    \item First, the appearance of several singular values $s_p$ of the $\mathbb{H}$ matrix whose magnitude is upper bounded by a factor, that we label $\Delta_{\mathrm{OBC}}$, and which decreases exponentially with system size, i.e., $s_p<\Delta_{\mathrm{OBC}}\propto e^{-N}$. We will use the calligraphic letter $\mathcal{N}_E$ to denote the set of these singular values.
    
    \item Second, that such singular values $s_p$ are separated from the bulk ones, $s_{n}\notin \mathcal{N}_E$, by a size-independent gap, that we label as $\Delta_{\mathrm{PBC}}$, i.e., $s_{n}\notin \mathcal{N}_E\geq\Delta_{\mathrm{PBC}}$ (see Appendix~\ref{sec:Deltas} for details). Note that $\Delta_\text{OBC}<\Delta_\text{PBC}$ is a necessary condition for the bulk-boundary condition to hold.
\end{itemize}

From the combination of both features and the steady-state decomposition that we write in Eq.~\eqref{eq:steady_state_coherences}, it is expected that the steady-state properties of the system are dominated by the "boundary" edge channels $s_p\in\mathcal{N}_E$ since their contribution to the steady-state is weighted by $s_p^{-1}\propto e^N$:
\begin{equation}
\langle b_j\rangle_\text{ss} \approx \sum_i\sum_{s_p\in\mathcal{N}_E}V_{j\ell}\frac{1}{s_p}U_{i\ell}^\star \Omega_i\;.\label{eq:steady_W}
\end{equation}

To check to which point these expectations are confirmed, we solve numerically the steady-state of the photonic lattice for a finite system and study two relevant magnitudes that provide information about the steady state features:
\begin{itemize}
\item One is the two-point Green's function defined by:
\begin{equation}
G_{ij}(\omega) = \left(\frac{1}{i\omega-i\mathbb{H}}\right)_{ij}\,,
\label{eq:green_function}
\end{equation}
that, at $\omega=0$ is directly related with the steady-state response of the system at a site $i$ to a coherent drive at site $j$ in the absence of parametric driving i.e. $g_s=0$ (see Appendix~\ref{sec:green_function} for details).

\item The other is the $k$-dependence structure of the steady-state coherences, i.e., $\mean{b_k}_\mathrm{ss}$, obtained by Fourier transforming Eq.~\eqref{eq:steady_state_coherences}. When $W\neq 0$, it is expected from Eq.~\eqref{eq:steady_W}, that such magnitude features a multi-peak structure that can be used as the smoking gun of multi-mode amplification, see Appendix~\ref{sec:fourier_multiexponential}.

\end{itemize}

\begin{figure}[t!]
    \includegraphics[width=0.99\columnwidth]{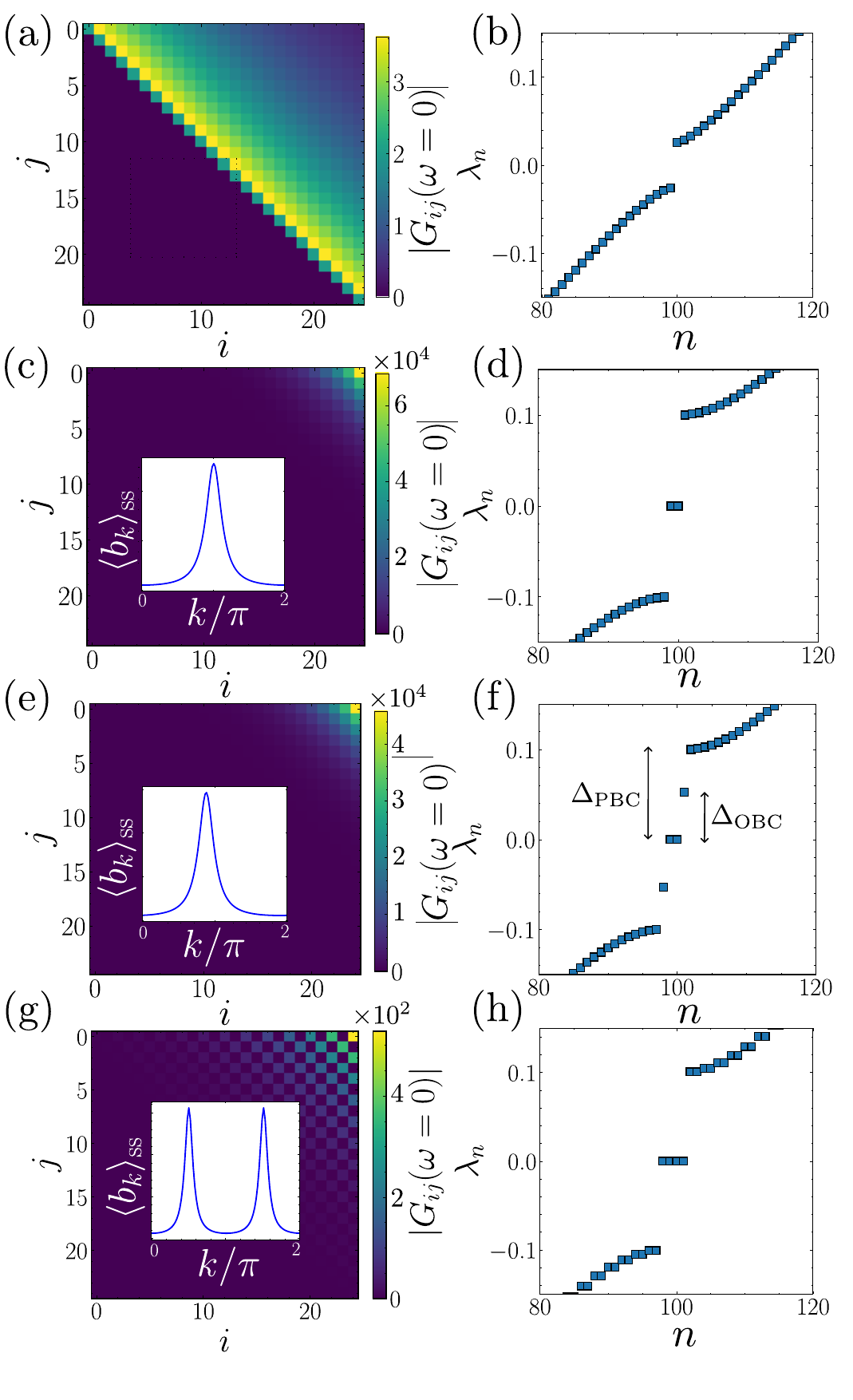}
    \caption{\textbf{Topological multi-mode amplification}. Amplification scenarios for $W=0$ (a, b), $W=1$ (c, d) $W=2$, and (f, g) $W=2$ in the particular case where $\Delta k\cdot a=\pi$ (g, h). (a, c, e, g) Absolute values of the spatial Green's function $|G_{ij}(\omega=0)|$. In the topologically non-trivial cases (c, e, g) we observe amplification, and an inset is added showing the spatial distribution of the steady state coherences $\langle b_r\rangle_\text{ss}$ when the system is coherently driven in the rightmost site with amplitude $\Omega_{i}=\Gamma\delta_{i,N}$. (b, d, f, g) Eigenvalues $\lambda_n$ of the doubled Hamiltonian $\mathcal{H}(\omega=0)$. In (f) we identify a gap in the bulk singular values $\Delta_\text{PBC}$ and see multiple eigenvalues within the gap, with maximum value $\Delta_\text{OBC}$. We are considering a waveguide with (a, b) a single-mode with $l_\kappa=/a=10$ and no driving, (c, d) single-mode with $l_\kappa=/a=10^3$ and $P/\Gamma=0.2$, (e, f) two modes with $(k_0a, k_1a)=(0,\pi/4)$, $l_\kappa/a=10^3$ and $P/\Gamma=0.2$, and (g, h) two modes with same parameters as (c, d), but $(k_0a,k_1a)=(0,\pi)$. The decay onto each mode is equal $\Gamma_\ell=\Gamma/n_\text{modes}$. We consider $N=100$ bosonic modes in all figures except for the Green's function distribution, where we take $N=25$ for visualization.}
  \label{fig:multimode_amplification}
\end{figure}

In Fig.~\ref{fig:multimode_amplification}, we plot these two magnitudes in the left column, together with the eigenvalue structure $\lambda_n$ of $\mathcal{H}$, of four illustrative situations appearing with the non-reciprocal multi-mode couplings of Eq.~\eqref{eq:chiral_multimode_self_energy}:
\begin{itemize}
    \item In Figs.~\ref{fig:multimode_amplification}(a-b), we consider a situation with non-reciprocal couplings, but with a trivial phase $W=0$. As expected, the eigenvalue structure, see panel (b), features a gap with no zero-energy eigenvalues. The two-point Green's function still features non-reciprocity, since  $G_{ij}(\omega=0)\neq 0$ only for $j>i$. However, the signal is not amplified and it rather decays as $|i-j|$ increases.

     \item In Figs.~\ref{fig:multimode_amplification}(c-d), we study a situation in which $W=1$, and thus the eigenvalue structure of $\mathcal{H}$ features a pair of zero-energy states $\pm s_p$, see Fig.~\ref{fig:multimode_amplification}(d). Consequently, the two-point Green's function is not only non-reciprocal, like in the previous case, but it also tends to exponentially accumulate at the end of lattice, a signature of the exponential amplification. The $k$-dependence of the steady-state coherences in this case also features a single peak, see inset in Fig.~\ref{fig:multimode_amplification}(c), as we expect from Eq.~\eqref{eq:steady_W}.

    \item In Figs.~\ref{fig:multimode_amplification}(e-f), we consider a situation with $W=2$. There, we observe in the eigenvalue structure that we obtain four different eigenvalues separated from the bulk modes. However, not all the eigenvalues are exponentially suppressed with the system size with the same scaling, and thus for a given system size, there is a gap between them bound by $\Delta_\mathrm{OBC}$. This has important implications in the steady-state behaviour since the smallest eigenvalue eventually dominates. This makes that eventually the two-point Green's function and the steady-state behaviour, see Fig.~\ref{fig:multimode_amplification}(e), shows virtually no difference with respect to the single-mode scenario of Fig.~\ref{fig:multimode_amplification}(c) for this small size limit.

    \item The multi-mode behaviour only manifests clearly in the steady-state when the zero-energy eigenvalues are truly degenerate. This can occur either in the $N\rightarrow\infty$ limit or because of some emergent symmetry. The latter case is what we consider in Figs.~\ref{fig:multimode_amplification}(g-h) for a situation with $W=2$ appearing for the special case $\Delta k \cdot a=\pi$. In that case, we indeed see in Fig.~\ref{fig:multimode_amplification}(h) that there are four eigenvalues $\lambda_p$ that are strictly equal, and consequently a two-peak structure appears in the $k$-coherences $\mean{b_k}_\mathrm{ss}$. This is a consequence of the particular structure of the couplings in this regime which we describe in Section~\ref{sec:effective_interactions} in which $J_{ij}\equiv 0$ and $\Gamma_{ij}$ is zero for the nearest-neighbouring sites. This leads to the oscillating structure appearing in the zero-frequency Green's function $|G_{ij}(\omega=0)|$ in which only the even sites are populated. More formally, one can say that such coupling structure leads to an exchange $\mathbb{Z}_2$-symmetry $U$ with the \emph{doubled Hamiltonian} such that $[U, \mathcal{H}]=0$. Hence, for every eigenstate $\ket{\psi}$ of $\mathcal{H}$, $U\ket{\psi}$ is another eigenstate with the same energy. This means that the winding number is even $W\in2\mathbb{Z}$, and the lowest singular values come in pairs.  
    
\end{itemize}

\subsection{Multi-mode, topological amplification: dynamics}
\label{sec:dynamics}

\begin{figure}[t!]
    \includegraphics[width=0.99\columnwidth]{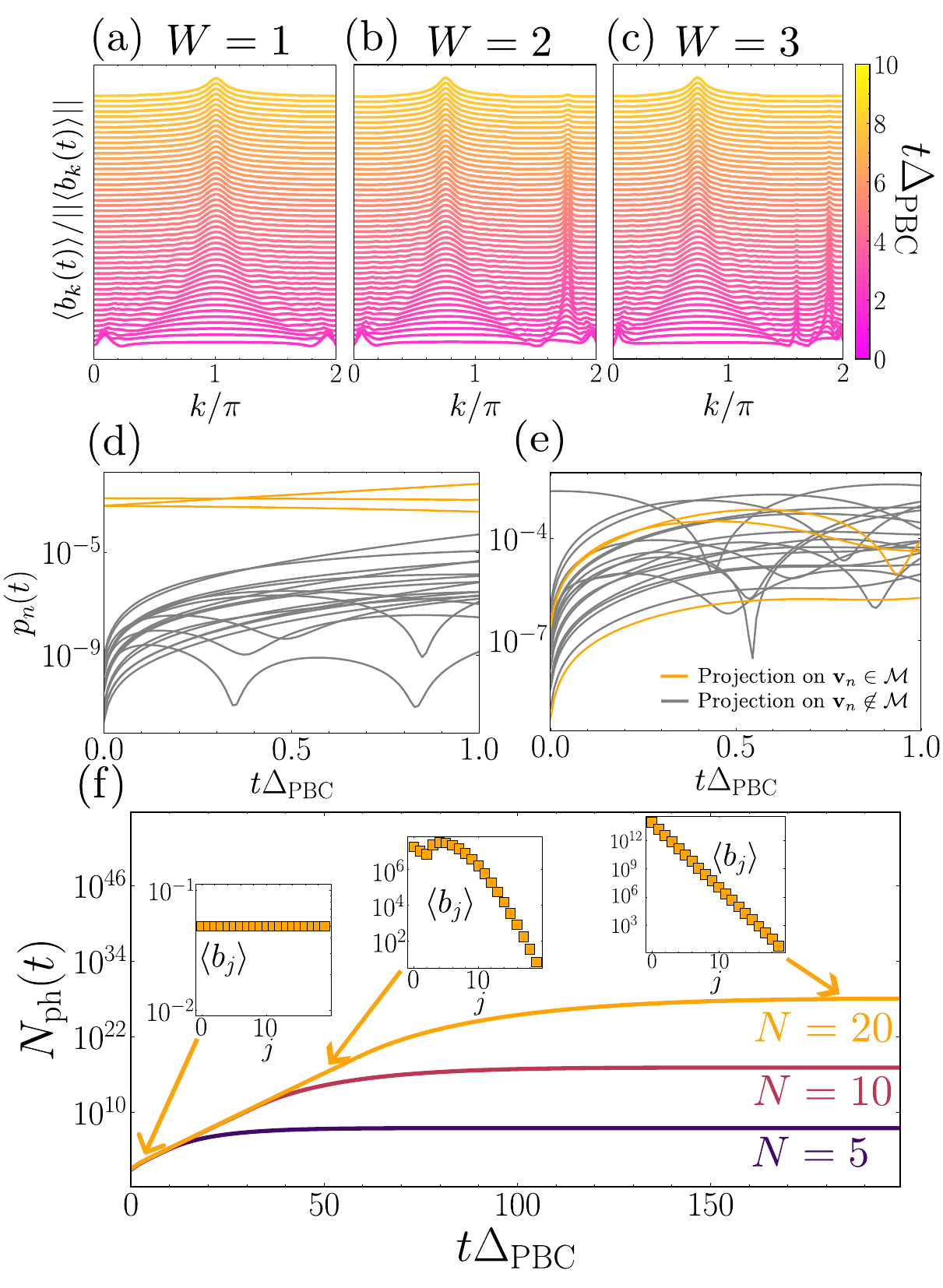}
    \caption{ \textbf{Topological dynamical metastability.} (a, b, c) Evolution of the normalized expected values $\langle b_k(t)\rangle$ as a function of time, for three situations with (a) $W=1$, (b) $W=2$ and (c) $W=3$ in the presence of a constant coherent drive in the rightmost local mode like in Fig.~\ref{fig:multimode_amplification}. The initial state correspond is uniformly distributed in real space $\langle b_r(t=0)\rangle=\langle b_{r'}(t=0)\rangle$ for all $r$, $r'$. In all cases, we observe an evolution to single-peaked distribution. However, at short times we observe a multi-peaked distribution of $\langle b_k\rangle$ for short times. (d, e) Dynamics of projections defined as Eq.~\eqref{eq:projection}, for (d) $\mathbf{b}(t=0)= (1/\sqrt{2})\mathbf{v}_{N}+(1/2)\mathbf{v}_{N-1}+(1/2)\mathbf{v}_{N-2}\in\mathcal{M}$ i.e. a linear combination of the three topological singular vectors, as $\mathcal{N}_E=\left\lbrace s_N, s_{N-1}, s_{N-2}\right\rbrace$, and (e) $\mathbf{b}(t=0)=\mathbf{v}_m \not\in\mathcal{M}$, i.e. $s_m\not\in\mathcal{N}_E$. We distinguish the evolution of the projections onto topological singular vectors that span $\mathcal{M}$ (in orange) and onto the rest (in gray). (f) Evolution of the total number of photons in the lattice $N_\text{ph}(t)$ in a topological phase for different system sizes $N$. In all cases, expected values of the field operators are amplified during a time that scales linearly on $N$, and follows stabilization at the steady state. Insets depict the bosonic modes coherences $\mathbf{b}(t)$, from an initial homogeneous distribution to an exponential localization at the amplifying edge. Simulation parameters: $P/\Gamma=0.7$, $l_\kappa/a=10^3$ and (a) single-mode and $W=1$ (b) two-mode with $(k_0a, k_1a)=(0, \pi/2)$ and $W=2$, (c-f) three-mode with $(k_0a,k_1a, k_2a)=(0,\pi/2,\pi/3)$ and $W=3$.}
    \label{fig:metastability}
\end{figure}

From the previous section, we conclude that the topological, multi-mode behaviour manifests in qualitatively different steady-state observables only under certain circumstances due to the different scaling of the \emph{zero}-energy eigenstates with system size, see Fig.~\ref{fig:multimode_amplification}(f). However, we will see in this section that precisely due to these different scalings, the transient behaviour features a non-trivial response, phenomenon that has been recently predicted in the general context of the theory of quadratic bosonic Lindbladians~\cite{ughrelidze2024interplay}.

To start illustrating it, we first study the time evolution of the $k$-dependent coherences $\mean{b_k(t)}$ starting from an initially spatially homogeneous configuration, i.e., $\mean{b_r(t=0)}=1/N$ for all $r$. This can be done using the formal solution of Eq.~\eqref{eq:eq_of_motion_b} which reads: 
\begin{align}
 \mathbf{b}(t) = e^{-i\mathbb{H}t} \left(\mathbf{b}_0-\mathbf{b}_\text{ss}\right)+\mathbf{b}_\text{ss}\,.\label{eq:solu}
\end{align}

In Fig.~\ref{fig:metastability}(a-c), we plot the evolution of $\mean{b_k(t)}$ in a finite system with $N=100$ for three different parameter regimes resulting in driven-dissipative topological phases with $W=1, 2$, and $3$, respectively. In all the cases, the $k$-dependent coherences end up stabilizing around a single peak structure, as expected from the behaviour described in Fig.~\ref{fig:multimode_amplification}(e). However, in the transient regime we do observe significant differences for the $W=2$ and $W=3$ situations with the appearance of additional peaks which decay much slower (in a timescale inversely proportional to $\Delta_\mathrm{OBC}$) than bulk modes, which decay in a time scale inversely proportional to $\Delta_\mathrm{PBC}$. 
Such dynamical multi-mode metastability is related to the appearance of additional topological amplification channels, 
which is why we label this phenomenon as topological dynamical multi-mode metastability.
This phenomenon is intimately connected with pseudospectrum theory in which the transient response of non-Hermitian systems is explained in terms of approximate eigenmodes, as explained in ~\cite{Flynn2023,ughrelidze2024}.

To formalize the connection with topological amplification, we can notice that in the Taylor expansion appearing in the solution of the coherence dynamics of Eq.~\eqref{eq:solu}:
\begin{align}
e^{-i\mathbb{H}t}\mathbf{b}=\left(\mathbf{1}-i t U SV^\dagger-\frac{t^2}{2}U SV^\dagger U SV^\dagger+...\right)  \mathbf{b}\;,
\end{align}
the expansion contains the identity $\mathbf{1}$ and a series of terms all ending in $V^\dagger$. Then, $e^{-i\mathbb{H}t}\mathbf{b}$ is equal to $\mathbf{b}$ plus a term that is proportional to $V^\dagger\mathbf{b}$. Let us define the orthonormal basis spanned by the vectors 
$\mathbf{v}_n$ formed by the $n^\text{th}$ column of $V$, i.e. $(\mathbf{v}_n)_i = V_{in}$. The vectors $\mathbf{v}_n$ such that $s_n\in\mathcal{N}_E$ will satisfy $V^\dagger \mathbf{v}_n= s_n\mathbf{v}_n\leq\Delta_\text{OBC}\mathbf{v}_n$, and therefore form a metastable subspace $\mathcal{M}$, since $e^{-i\mathbb{H}t}\mathbf{v}_n$ will be equal to $\mathbf{v}_n$ plus an exponentially suppressed term. On the contrary, vectors $\mathbf{v}_n$ with $s_n\not\in\mathcal{N}_E$ fulfill that
$V^\dagger \mathbf{v}_n \geq\Delta_{\rm PBC} \mathbf{v}_n$ and they evolve on a time scale of the order of $1/\Delta_{\rm PBC}$. Since $\Delta_\text{PBC}$ is size-independent and $\Delta_\text{OBC}$ is exponentially suppressed with $N$, it is expected that vectors in $\mathcal{M}$ evolve significantly slower than the ones outside $\mathcal{M}$.

A way to evidence the metastable subspace is to keep track of the projections of the evolving vector of coherences $\mathbf{b}$ with the orthonormal basis defined by the columns of $V$

\begin{equation}
p_n(t) = \left|\langle \mathbf{b}(t)|\mathbf{v}_n\rangle \right|^2    \;,
\label{eq:projection}
\end{equation}
which, due to the non-Hermitian nature of the evolution, are not necessarily upper bound by 1. In Fig.~\ref{fig:metastability}(d-e), we plot the time evolution of the projections for a parameter regime situation with $W=3$ distinguishing the contribution of the topological singular vectors (in orange) and bulk ones (in gray), respectively. In particular, in Fig.~\ref{fig:metastability}(d) we consider a situation where the initial state belongs to the metastable sector, i.e., $\mathbf{b}(t=0)\in\mathcal{M}$, showing how the metastable states projections are not altered for a time scale of the order of $1/\Delta_\text{PBC}$. In Fig.~\ref{fig:metastability}(e) we start with the opposite situation, that is, with an initial bulk vector state $\mathbf{b}(t=0)=\mathbf{v}_m\notin\mathcal{M}$, whose projection $p_m(t)$ evolves much faster, since the associated singular value $s_m$ is larger than the topological ones in $\mathcal{N}_E$.

Up to now, we have focused on the qualitative differences of the driven-dissipative topological phases but have not discussed the dynamical amplification behaviour itself. To illustrate the amplification dynamics, we calculate the time evolution of the total number of coherent photons in the lattice, i.e., $N_\mathrm{ph}(t)=\sum_r|\mean{b_r(t)}|^2$, starting by an initially homogeneous situation, i.e., $\mean{b_r(t=0)}=1/N$, with $N$ being the number of lattice sites. In Fig.~\ref{fig:metastability}(f), we plot $N_\mathrm{ph}(t)$ for different system sizes $N=5, 10$ and $20$ in the different colors depicted in the legend. All of them share a similar behaviour, that is, the photon population increases until it stabilizes to its steady-state value. The duration of this transient dynamics scales linearly on $N$, because it is directly related to the time required to the initial signal to travel along the chain and accumulate at the amplifying edge, as represented in the insets of Fig.~\ref{fig:metastability}(f). Another signature of the topological amplification that is clear from Fig.~\ref{fig:metastability}(f) is that the photon number in the steady-state $N_\mathrm{ph}(t\rightarrow\infty)$ increases exponentially with the system size (note the logarithmic scale of the figure). 

Such exponential increase in the transient stage is characteristic of unstable systems and can also be explained under the light of
pseudospectral theory \cite{ughrelidze2024,Flynn2023}. 
The latter permits to characterize the time-evolution of non-Hermitian systems in terms of approximate unstable eigenmodes, leading to an an initial evolution stage of exponential amplification. At long times, exact eigenmodes take over and lead to the stable steady-state.

Let us finally remark that in real experimental situations there will always be other mechanisms, e.g., interactions, that will renormalize such exponential amplification. In fact, this is an interesting research direction to consider after this work.

\section{Potential implementation}
\label{sec:implementation}

The key ingredient of the phenomena that we have explored along this manuscript is the engineering of the non-reciprocal, multi-mode, long-range hoppings of Eq.~\eqref{eq:chiral_multimode_self_energy}. As we explain in Section~\ref{sec:effective_interactions}, these couplings can emerge by either:
\begin{itemize}
    \item Coupling the cavity modes to a one-dimensional channel hosting chiral, multiple waveguide modes, i.e., propagating only into one direction.

    \item Assuming the one-dimensional channels host multiple bidirectional waveguide modes, but that the cavities only couple to the ones propagating in one direction, e.g., using optical spin-orbit coupling~\cite{lodahl17a,Bliokh2015}.
\end{itemize}

\begin{figure}[t!]
    \includegraphics[width=0.99\columnwidth]{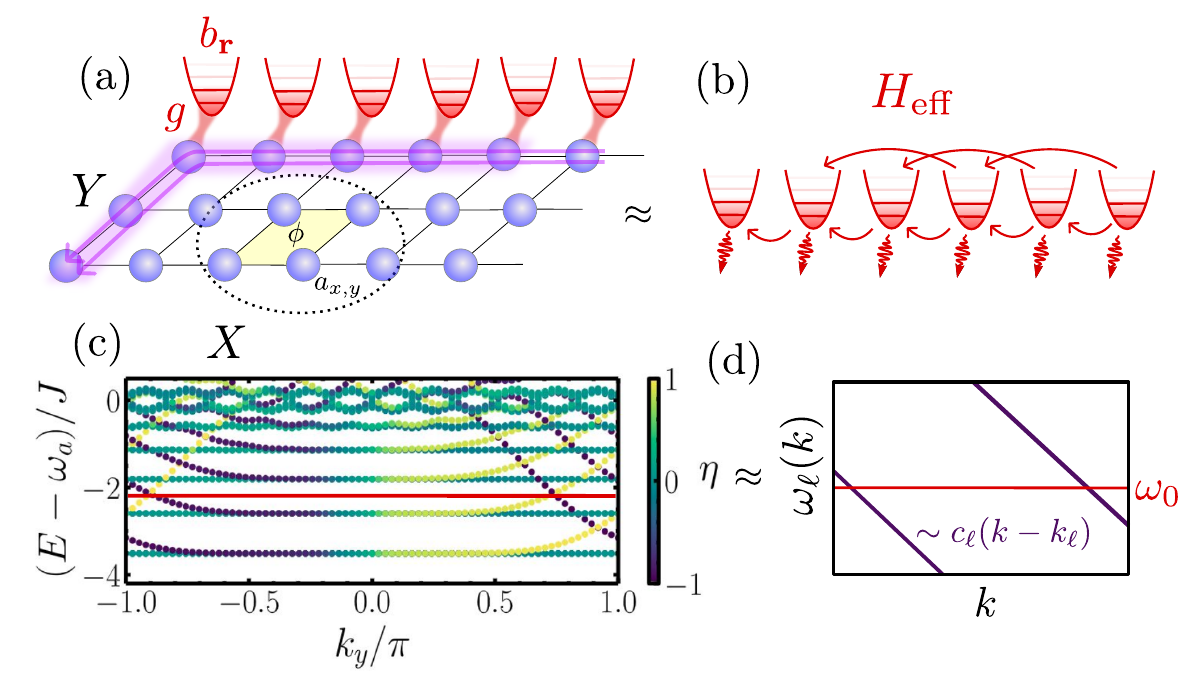}
    \caption{\textbf{Scheme of a potential experimental implementation of the non-reciprocal couplings.} (a) Red parabolas with harmonic levels depict the array of resonators representing the local bosonic modes $b_r$. These resonators are coupled to the edge of a lattice of superconducting resonators, depicted as purple globes, representing the Harper-Hofstadter lattice. The dotted circle encoloses a lattice plaquette, coloured in yellow. A photon circulating such plaquette will catch a global Aharonov-Bohm-like phase $e^{i\phi}$. (b) After adiabatic elimination of the lattice photons, that are topological edge photons at the frequency of the resonators, the local bosonic modes become effectively coupled following the effective Hamiltonian in Eq.~\eqref{eq:effective_Hamiltonian}. (c) Lowest part of the spectrum of a Harper-Hofstadter lattice $\phi=1/9$ with periodic boundary conditions along $Y$ and open along $X$. Each eigenenergy is plotted with a color index $\eta=-1+\sum_{x}(2x/L-1)|\Psi(x)|^2$, denoting the localization at each $X$-edge, which is directly related to the chirality of the modes. (d) In the Markovian regime, the only relevant bath modes are those close to resonance to $\omega_0$, which energy can be approximated e.g. by a linearization $\omega_\ell(k)\sim c_\ell(k-k_\ell)$.}
    \label{fig:Implementation}
\end{figure}

In this section, we discuss a potential physical implementation based on the first idea. The idea consists in coupling the $b_i$ cavity modes to the boundary of a two-dimensional Chern insulator hosting multiple-edge modes~\cite{Vega2023TopologicalQED,Skirlo2014,Skirlo2015}, see Figs.~\ref{fig:Implementation}(a-b) for an schematic representation of the setup. In particular, this can be achieved in a two-dimensional cavity array with complex tunnelings emulating an artificial magnetic field, i.e., the so-called Harper-Hofstadter lattice model~\cite{hofstadter76a}. This model is described by a Hamiltonian:
\begin{equation}
H = -J\sum_{x,y}a_{x+1,y}^\dagger a_{x,y}+e^{i2\pi\phi x}a_{x,y+1}^\dagger a_{x,y}+\text{H.c.}    \;.
\end{equation}
with $a_{x,y}$ being the bosonic operator associated to the cavity at the $(x,y)$ position. As shown in Ref.~\cite{Vega2023TopologicalQED}, if $\phi=1/q$ with $q\in\mathbb{N}$, the spectrum of the Harper-Hofstadter Hamiltonian with open boundary conditions in one spatial direction (X) and periodic in the other (Y) features a series of flat bands, and $n$ resonant propagating edge modes in the $n^\text{th}$ band-gap. We introduce a color scale through a parameter $\eta$, bounded between $-1$ and $1$, which defines the edge localization of the eigenstates in the X direction. This localization is directly connected to chirality: edge modes localized at $\eta=-1$ are right-moving, while those at $\eta=1$ are left-moving. We see that if we place a cavity array with frequencies $\omega_c$ within the second band-gap at $\eta=-1$ localization, as indicated with the red line of Fig.~\ref{fig:Implementation}(c), they will be resonant to two chiral modes with energy dispersion $\omega_\ell(k)$ as depicted in Fig.~\ref{fig:Implementation}(d). Thus, assuming that the coupling is weak enough, the multiple edge channels will result in the $\Gamma_{ij},J_{ij}$ derived in Section~\ref{sec:effective_interactions}.

There are several platforms in which such Harper-Hofstadter bosonic lattice has been experimentally realized. One of the most recent ones is coupled microwave resonators~\cite{Owens2022}, in which by coupling additional resonators our model can be naturally implemented. Other potential platform are bosonic ultra-cold atoms in optical lattices, in which the Harper-Hofstadter Hamiltonian has already been engineered~\cite{aidelsburger2015measuring}. By harnessing the recent advances in state-dependent optical lattices~\cite{krinner18a,Kwon2022FormationLattice,snigirev2017towards,Heinz2020}, one can simulate the coupling to additional modes and emulate our models.

\section{Conclusions $\&$ Outlook}

Summing up, in this work we unveil a different class of non-reciprocal, long-range couplings emerging from the chiral exchange of excitations via a common, multi-mode, one dimensional bath. Despite their perfect non-reciprocity, such couplings also depend critically of their separation, something not possible with standard chiral single-mode channels. By considering different excitation mechanisms, we demonstrate that these couplings stabilize different topological amplification phases with multiple channels, and characterize their steady-state and dynamical features, finding a metastable regime linked to existence the multiple amplifying channels. A natural outlook of this work is to consider some type of interaction in the photonic lattice and study how they affect the steady-state and dynamical behaviour of these lattices.

\begin{acknowledgments}
The authors acknowledge support from the Proyecto Sin\'ergico CAM 2020 Y2020/TCS-6545 (NanoQuCo-CM), the CSIC Research Platform on Quantum Technologies PTI-001 and from Spanish projects PID2021-127968NB-I00 funded by MICIU/AEI/10.13039/501100011033/ and by FEDER Una manera de hacer Europa, and TED2021-130552B-C22 funded by  MICIU/AEI /10.13039/501100011033 and by the European Union NextGenerationEU/ PRTR, respectively. On top of that, AGT acknowledges  a 2022 Leonardo Grant for Researchers and Cultural Creators, and BBVA Foundation.
AMH acknowledges support from Fundaci\'{o}n General CSIC's ComFuturo programme which has received funding from the European Union's Horizon 2020 research and innovation programme under the Marie Sk\l{}odowska-Curie grant agreement No. 101034263. 
\end{acknowledgments}

\bibliographystyle{plain}
\bibliography{references_doi,referencesAlex}

\onecolumn\newpage
\appendix
\appendix

\section{Derivation of the chiral multi-mode master equation}
\label{sec:master_equation}
Along this work, we consider a set of local cavity modes as an open quantum system coupled to a waveguide environment or bath. In the conditions of the Born-Markov approximation, the adiabatic elimination of the waveguide photons yields a master equation description of the system dynamics described by Eq.~\eqref{eq:meq} of the main text, where the effective Hamiltonian describing the coupling between cavities reads:
\begin{equation}
H_\text{eff}=\sum_{ij}\left(J_{ij}-i\frac{\Gamma_{ij}}{2}\right)b_i^\dagger b_j\;.    
\end{equation}

Here, we will show that, for a chiral multi-mode waveguide bath, the effective coherent ($J_{ij}$) and incoherent ($\Gamma_{ij}$) couplings between the cavity modes are given by Eq.~\eqref{eq:chiral_multimode_self_energy} of the main text.\\ 

To show this, we will compute the collective \emph{self-energy} $\Sigma_{ij}(z)$ of the cavity modes, a complex function that accounts for the interaction with its environment~\cite{CohenTannoudji1998}. The couplings $J_{ij}$ and $\Gamma_{ij}$ are determined by the real and imaginary parts of this function evaluated at the frequency of the cavity modes $\omega_c$:
\begin{equation}
J_{ij}-i\frac{\Gamma_{ij}}{2} = \Sigma_{ij}(\omega_c+i0^+)\;,    
\end{equation}
where the $i0^+$ sets an integration prescription to ensure convergence. We can write the interaction Hamiltonian between the cavity modes and the multi-mode waveguide as $H_\text{int}=\sum_i b_i B_{r_i}^\dagger+\text{H.c.}$, where $B_{r_i}$ is an collective waveguide operator defined as:
\begin{equation}
B_{r_i}^\dagger = \frac{1}{2\pi}\sum_\ell\int dk\; g_k e^{ikr_i}A_{\ell, k}^\dagger\;,   
\end{equation}

Using these definitions, the self-energy of the bosonic modes can be shown to be given by~\cite{breuer2002theory}:
\begin{equation}
\Sigma_{ij}(\omega_c+i0^+)=\int_0^\infty d\tau \langle B_{r_i}^\dagger (\tau)B_{r_j}(0)\rangle e^{i(\omega_c+i0^+)\tau}\;.   
\end{equation}

Assuming: (i) the bath eigenvalues are of the form $\psi^\ell_k(r)=(1/\sqrt{N}) e^{ikr}$, labeled by a mode index $\ell$ and a momentum $k$; (ii) $H_B\psi^\ell_k(r)=\omega_\ell(k)\psi_k(r)$; (iii) and that the waveguide modes are in the vacuum state, one arrives to

\begin{equation}
\Sigma_{ij}(\omega_c+i0^+)=\frac{1}{2\pi}\sum_\ell\int dk\;\frac{|g_k|^2\; e^{ik(r_i-r_j)}}{\omega_c+i0^+-\omega_\ell(k)}    \;.
\end{equation}

To continue with the calculation, we need to impose the chirality of the transfer, that can either occur either because $\omega_\ell(k)$ only exists for $k\lessgtr 0$, or because only right- or left-moving photons are coupled to the cavity modes i.e. $|g_k|^2\propto \theta(\pm k)$, where $\theta$ is the Heaviside step function. In either case, one can account for both behaviours by restricting the $k$-integral to either positive or negative momenta. For concreteness, we will do it for positive one:
\begin{equation}
\Sigma_{ij}(\omega_c+i0^+)=\frac{1}{2\pi}\sum_\ell\int_0^\infty dk\;\frac{|g_k|^2\; e^{ik(r_i-r_j)}}{\omega_c+i0^+-\omega_\ell(k)}    \;.
\end{equation}
where we have also extended the range of integration to infinity. The latter is a good approximation in the Markovian regime where the shape of the integral will be dominated by the momenta resonant to $\omega_c$, that we denote by $k_\ell$, i.e., $\omega_\ell(k_\ell)=\omega_c$. Now, using that assumption and applying Sokhotski-Plemelj theorem, we arrive to 
\begin{align}
\Sigma_{ij}(\omega_c+i0^+)&=\sum_\ell \frac{1}{2\pi}\int_{0}^\infty dk\;|g_k|^2\;e^{ik(r_i-r_j)}\left(-i\pi\delta(\omega_c-\omega_\ell(k))+\text{P}\frac{1}{\omega_c-\omega_\ell(k)}\right)\\
&\approx \sum_\ell -i\frac{|g_{k_\ell}|^2}{2v_\ell}\left(e^{ik_\ell(r_i-r_j)}+\frac{1}{2\pi}\text{P}\int_0^\infty dk\;\frac{e^{ik(r_i-r_j)}}{k-k_\ell}\right)\;.
\end{align}
where we introduce the notation $v_\ell(k)\equiv \partial_{k}\omega_\ell(k)$, for the group velocity, P for the Cauchy principal value, and the only implicit assumption is that $\omega_c\in \omega_\ell(k)$. To continue with the calculation, we change the variables $k-k_\ell=q$ and use the identity:
\begin{equation}
\text{P}\int_{-\infty}^\infty dx\;\frac{e^{iax}}{x}=i\pi\text{sign}(a)\,,
\end{equation}
to arrive to:
\begin{equation}
\Sigma_{ij}(\omega_c+i0^+)=\sum_\ell -i\frac{g_{k_\ell}^2}{2v_{\ell}}e^{ik_\ell(r_i-r_j))}\left(1+\text{sign}(r_i-r_j)\right)    \;.
\end{equation}
Now, by introducing the Markovian decay rate into the $\ell^\text{th}$ waveguide mode $\Gamma_\ell\equiv g_{k_\ell}^2/v_\ell$, we finally get 
\begin{equation}
\Sigma_{ij}(\omega_c+i0^+)=\sum_\ell -i\frac{\Gamma_\ell}{2}e^{ik_\ell(r_i-r_j))}\left(1+\text{sign}(r_i-r_j)\right)    \;.
\end{equation}

The structure of the self-energy offers a clear intuition on the chiral character of the modes: being proportional to $1+\text{sign}(r_i-r_j)$. Thus, it vanishes if $r_i<r_j$ but does not if $r_i\geq r_j$. This expression holds in the ideal situation where photon losses are negligible.

To generalize this self-energy to the lossy case, we introduce the propagation length $l_\kappa$, an average distance photons travel before being dissipated to the environment, that provides an overall exponential decay as
\begin{align}
\Sigma_{ij}(\omega_c+i0^+)=\sum_\ell -i&\frac{\Gamma_\ell}{2}e^{ik_\ell(r_i-r_j))-|r_i-r_j|/l_\kappa}\left(1+\text{sign}(r_i-r_j)\right)    \;.
\end{align}

\section{On the connection between eigenvalues of $\mathcal{H}$ and singular values of $\mathbb{H}$}
\label{sec:eigvals_and_singular_values}

As shown in Ref.~\cite{Porras2019, Ramos2021}, the eigenvalues of the doubled Hamiltonian $\mathcal{H}$ defined in Eq.~\eqref{eq:doubled_H} come in pairs of positive and negative singular values of the dynamical matrix $\mathbb{H}$ i.e. $\lambda_n=\pm s_n$. Due to the importance of this result in our work, we revisit here its proof. Let us start by noticing that the doubled Hamiltonian $\mathcal{H}$ can be written in terms of the singular value decomposition $\mathbb{H}=USV$ as:
\begin{equation}
\mathcal{H} = 
\begin{pmatrix}
0 & \mathbb{H} \\
\mathbb{H}^\dagger & 0
\end{pmatrix}
=
\begin{pmatrix}
0 & USV^\dagger \\
VSU^\dagger & 0
\end{pmatrix}\;.
\end{equation}
Using that $U$ and $V$ are unitary matrices i.e. $V^\dagger V=U^\dagger U=\mathbf{1}$, it is straight-forward to verify that $\mathcal{H}$ can be diagonalized as
\begin{equation}
\mathcal{H} = A^\dagger 
\begin{pmatrix}
S & 0\\
0 & -S
\label{eq:singular_value_diagonalization}
\end{pmatrix}
A
\end{equation}
where the diagonal basis transformation $A$ is given by
\begin{equation}
A = 
\begin{pmatrix}
U & U \\
V & -V
\end{pmatrix}\;.
\end{equation}
From the diagonal matrix in Eq.~\eqref{eq:singular_value_diagonalization}, it follows that the eigenvalues of $\mathcal{H}$ come in pairs consisting in the singular values of $\mathbb{H}$ with the positive and negative sign.\\

\section{On the winding number of the doubled Hamiltonian}
\label{sec:winding_number}

The driven-dissipative topological phases we explore in this work are characterized by the winding number of the doubled Hamiltonian defined by:
\begin{equation}
W = \frac{1}{4\pi i}\int_\text{BZ}dk\;\text{Tr}\left(\tau_z\mathcal{H}(k)^{-1} \partial_k\mathcal{H}(k)\right)\;,    
\end{equation}
where $\tau_z$ is a matrix representation of the chiral symmetry and assumed a doubling of the degrees of freedom. In this section, we derive a simplified expression for $W$ in the absence of parametric driving $g_s=0$. In that case, the doubled Hamiltonian can be simply written as
\begin{equation}
\mathcal{H}(k)=
\begin{pmatrix}
0 & h(k)\\
h^\star(k) & 0
\end{pmatrix}
\end{equation}
If we now plug the doubled Hamiltonian in Eq.~\eqref{eq:winding_number_matrix}, we get 
\begin{align}
W =& \frac{1}{4\pi i}\int_\text{BZ}dk\;\text{Tr}\left[\tau_z\begin{pmatrix}
0 & h^\star(k)^{-1}\\
h(k)^{-1} & 0
\end{pmatrix}
\begin{pmatrix}
0 & \partial_k h(k)\\
\partial_k h^\star(k) & 0
\end{pmatrix}\right] \nonumber\\
=&
\frac{1}{4\pi i}\int_\text{BZ}dk\;\text{Tr}\left[\tau_z\begin{pmatrix}
\partial_k \log h^\star(k) & 0\\
0 & \partial_k \log h(k)
\end{pmatrix}
\right]  \;.
\end{align}
The $\tau_z$ term is a matrix representation of the chiral symmetry of the doubled Hamiltonian i.e. $\tau_z\mathcal{H}\tau_z=-\mathcal{H}$. We will take $\tau_z=\begin{pmatrix}
    -1 & 0\\
    0 & 1
\end{pmatrix}=-\sigma_z$. Noting that only the imaginary part of the integral is non-zero, and having that $\text{Im}(\log x^\star)=-\text{Im}(\log x)$, we find that
\begin{align}
W &= \frac{1}{4\pi i}\int_\text{BZ}dk\;2\partial_k\log h(k)=\frac{1}{2\pi i}\int_\text{BZ}dk\;\partial_k\log h(k) \;,
\end{align}
which is the simplified expression in Eq.~\eqref{eq:winding_number_scalar}.\\

Further, we can show that, still in the $g_s=0$ regime, the winding number is upper bounded by the number of waveguide modes. To prove this, let us perform the change of variable $y=e^{ik}$ in the expression for the winding number, replacing $h(k)$ by the non-Hermitian spectrum of our chiral multi-mode waveguide (see Eq.~\eqref{eq:nonHermitian_spectrum}):
\begin{align}
W=&\frac{1}{2\pi i}\int_\text{BZ}dk\; \partial_k \log\left(i\frac{P}{2}-i\sum_\ell \left[\frac{\Gamma_\ell}{2}+\frac{\Gamma_\ell e^{i(k+k_\ell)-1/l_\kappa}}{1-e^{i(k+k_\ell)-1/l_\kappa}}\right]\right)\nonumber\\
=& -\frac{1}{2\pi}\int_{\mathbb{S}^1}\frac{dy}{y}\;\partial_y\log\left(i\frac{P}{2}-i\sum_\ell \left[\frac{\Gamma_\ell}{2}+\frac{\Gamma_\ell ye^{ik_\ell-1/l_\kappa}}{1-ye^{ik_\ell-1/l_\kappa}}\right]\right)\;.
\end{align}
Now, let $F(y)$ be the argument of the logarithm. The winding number will be equal to the number of solutions of $F(y)=0$ that lie within the unit circle of the complex plane, $\mathbb{S}^1$. It can be directly shown than the equation can be recast into a polynomial equation of degree equal to $n_\text{modes}$, and therefore the number of its solution within $\mathbb{S}^1$ is upper bounded by the total number of solutions in the complex plane, which is $n_\text{modes}$. Therefore, we conclude that $W\leq n_\text{modes}$.

\section{Diagonalization of the non-Hermitian Hamiltonian}
\label{sec:non_Hermitian_Hamiltonian_diagonalization}
Here, we present the diagonalization of the dynamical matrix $\mathbb{H}$, presented in Eq.~\eqref{eq:nonHermitian_spectrum} of the main text. Let us start by considering a general traslationally invariant Hamiltonian in a one-dimensional lattice, which can be written as follows
\begin{equation}
\mathbb{H} = \sum_r\sum_{n\in\Lambda} C_n \;b_r^\dagger  b_{r+n} \;.
\end{equation}

Here, $r$ is a coordinate summed over the whole lattice, while the sum in $n$ is performed over a set neighbouring sites $\Lambda$ of $r$ that are coupled to $b_r$. This Hamiltonian is manifestly invariant under translations, since the coupling $C_n$ depends only in $n$ and not in $r$, and can be readily diagonalized as 
\begin{equation}
h(k) = \sum_{n\in\Lambda} C_n e^{-ikna}    \;,
\end{equation}
where $a\equiv |r_{j+1}-r_j|$. Introducing the expression for the couplings $C_n$ corresponding to our effective Hamiltonian in Eq.~\eqref{eq:effective_Hamiltonian} of the main text, we get 
\begin{equation}
h(k) = \frac{P}{2}-i\sum_{n}\sum_\ell \frac{\Gamma_\ell}{2} e^{-ikna}e^{ik_\ell n}e^{-|n|a/l_\kappa}\left(1+\text{sign}(n)\right)    
\end{equation}

Now, due to the $1+\text{sign}(n)$ factor, only terms with $n\geq 0$ will be different from zero, which is directly linked to the fact that $\Lambda=\mathbb{N}$ in our problem. Then, we can rewrite $h(k)$ as
\begin{equation}
h(k) = \frac{P-\Gamma}{2}-i\sum_{n=1}^\infty\sum_\ell \Gamma_\ell e^{-ikna}e^{ik_\ell na}e^{-na/l_\kappa}   \;,
\end{equation}
where $\Gamma\equiv \sum_\ell \Gamma_\ell$. Now, we notice that the sum in $n$ is a geometric series, that is convergent as long as $l_\kappa>0$ i.e. if the couplings are finite-range. Performing the sum yields
\begin{equation}
h(k) = i\frac{P-\Gamma}{2}-i\sum_\ell\frac{\Gamma_\ell e^{i(k_\ell-k)a-a/l_\kappa}}{1-e^{i(k_\ell-k)a-a/l_\kappa}}\;,
\end{equation}
which is the Eq.~\eqref{eq:nonHermitian_spectrum} of the main text.

\section{On the Green's function}
\label{sec:green_function}

In the main text, we claim that the Green's function defined in Eq.~\eqref{eq:green_function} is connected to the linear response of the system at a site $i$ to a coherent driving in site $j$. Here, we will justify this affirmation in the absence of parametric driving i.e. $g_s=0$. Let us consider the equation of motion for the local cavity modes coherences in this regime:
\begin{equation}
\frac{d\mathbf{b}(t)}{dt} = -i\mathbb{H}\mathbf{b}(t) + i\mathbf{\Omega}(t)\;. 
\end{equation}
This is a linear first-order differential equation, and as such can be solved by Fourier transform. In terms of the Fourier transform variables, defined as
\begin{align}
\mathbf{b}(\omega) =& \frac{1}{2\pi}\int_{-\infty}^\infty
 dt\; e^{-i\omega t}\mathbf{b}(t)\\
\mathbf{\Omega}(\omega) =& \frac{1}{2\pi}\int_{-\infty}^\infty
 dt\; e^{-i\omega t}\mathbf{\Omega}(t)\;,
\end{align}
the equation of motion can be rewritten as
\begin{equation}
-i\omega\mathbf{b}(\omega) = -i\mathbb{H}\mathbf{b}(\omega)+i\mathbf{\Omega}(\omega)  \;.
\end{equation}
We then conclude that its solution is given by 
\begin{equation}
\mathbf{b}(\omega) = -\left(\frac{1}{i\omega-i\mathbb{H}}\right)\mathbf{\Omega}(\omega) \equiv - G(\omega)\mathbf{\Omega}(\omega)\;.
\end{equation}
Therefore, the response at a site $i$ of the system to a driving at a site $j$ will be determined by $G_{ij}(\omega)$, as claimed. In the steady-state, this response will be given by $G_{ij}(\omega=0)$, as stated in Eq.~\eqref{eq:ss_solution} of the main text.

\section{Scaling of $\Delta_\text{PBC}$ and $\Delta_\text{OBC}$ with the system size}
\label{sec:Deltas}

\begin{figure}[h!]
    \centering
    \includegraphics[width=0.59\columnwidth]{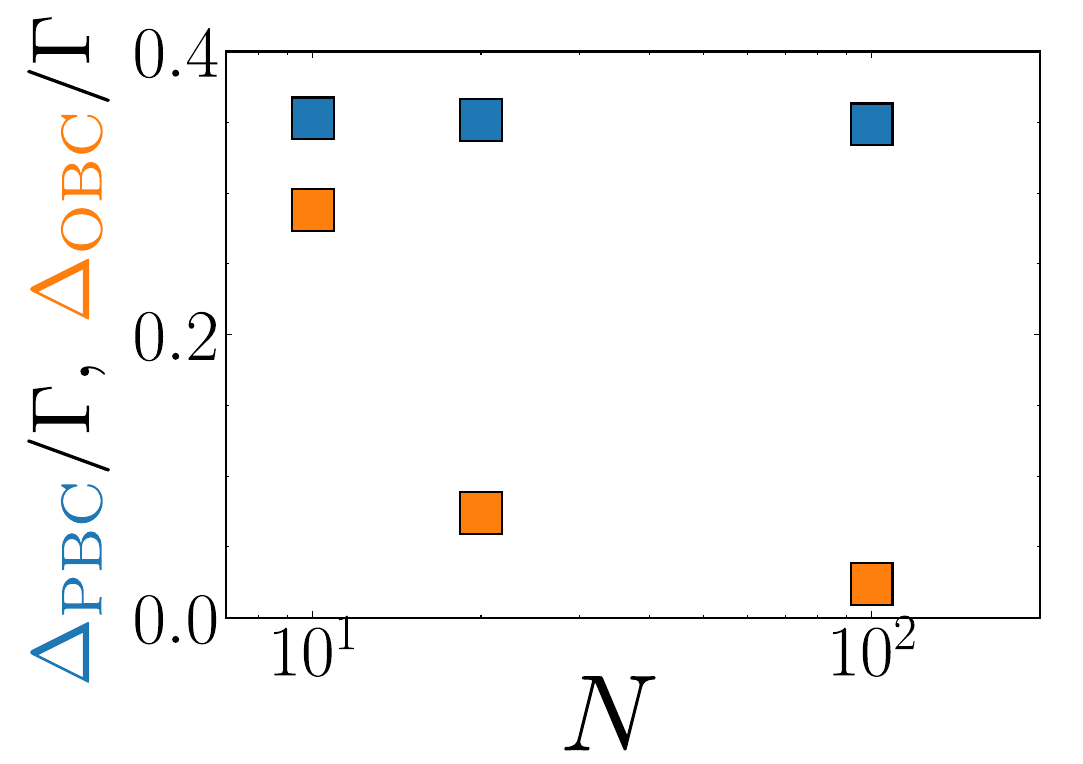}
    \caption{Scaling of the singular values gaps $\Delta_\text{PBC}$ (in blue) and $\Delta_\text{OBC}$ (in orange), plotted in units of $\Gamma$. Simulation parameters:$(k_0a, k_1a,k_2a)=(0,\pi/2,\pi/3)$, $P/\Gamma=0.7$ and $l_\kappa/a=10^3$, with no parametric driving i.e. $g_s=0$.}
    \label{fig:Deltas}
\end{figure}

In the main text, we claimed that $\Delta_\text{PBC}$ is a size-independent magnitude, while $\Delta_\text{OBC}$ decreases exponentially as the system size $N$ increases. This behaviour is expected from the equivalence between the eigenvalues of the doubled Hamiltonian $\mathcal{H}$ and the singular values of the dynamical matrix $\mathbb{H}$ stated in Eq.~\eqref{eq:eig}: since $\Delta_\text{OBC}$ can be seen as the energy of a topological edge state, it is expected to decay exponentially with the system size $N$~\cite{asboth15}, while $\Delta_\text{PBC}$ determines the band-gap of the doubled Hamiltonian $\mathcal{H}$, that is size-independent. In this Appendix, we present numerical simulations that justify this claim. In Fig.~\ref{fig:Deltas} we show the values of these magnitudes for three values of $N$, and observe the decay of $\Delta_\text{OBC}$ as $N$ increases, while the value of $\Delta_\text{PBC}$ remains unaltered. The bulk-boundary correspondence is well defined as long as $\Delta_\text{OBC} < \Delta_\text{PBC}$. The condition $\Delta_\text{OBC} \approx \Delta_\text{PBC}$ implies the merging of the topological edge state with the bulk and, thus, the breaking of the bulk-edge correspondence.

\section{On the Fourier transform of a multi-exponential function}
\label{sec:fourier_multiexponential}

In Sec.~\ref{sec:multi_mode_amplification} of the main text, we claim that $\langle b_k\rangle$, the discrete Fourier transform of the vector of cavity mode coherences, is an insightful probe of the multi-mode structure of the steady state. Here, we will show why. As we discussed in the main text, the spatial distribution of the steady-state coherences are domintated by topological \emph{zero} singular values $\mathcal{N}_E$: 

\begin{equation}
\langle b_j\rangle_\text{ss} \approx \sum_i\sum_{s_p\in\mathcal{N}_E}V_{j\ell}\frac{1}{s_p}U_{i\ell}^\star \Omega_i\;.
\end{equation}

Following the notation we use in Sec.~\ref{sec:dynamics}, letting $\mathbf{v}_n$ the vector obtained by taking the $n^\text{th}$ column of $V$, this equation tells us that $\mathbf{b}_\text{ss}$ is a linear combination of the vectors $\mathbf{v}_{p}$ associated to $s_p\in\mathcal{N}_E$. These vectors are exponentially localized at the amplifying edge. Therefore, $\mathbf{b}_\text{ss}$ is a sum of exponential contributions, each of them coming from a waveguide mode. Let us now show that the discrete Fourier transform of a multi-exponential function is an useful way of retrieving the number of contributing terms. Let $f:\left\lbrace 0,1,...,N-1\right\rbrace \rightarrow \mathbb{C}$ be a complex function depending on a non-negative integer index $j$. We say that $f(j)$ is multi-exponential if it is of the form
\begin{equation}
f(j) = \sum_{\ell} c_\ell e^{(i\varphi_\ell-1/\lambda_\ell)j}    
\end{equation}
The discrete Fourier transform of $f$ reads
\begin{equation}
f(k)=\frac{1}{\sqrt{N}}\sum_r e^{-ikr}f(r)\approx \frac{1}{\sqrt{N}}\sum_\ell \frac{c_\ell}{1-e^{i(\varphi_\ell-k)-1/\lambda_\ell}} \;,
\end{equation}
where we assumed that $Na\gg \lambda_\ell$. The function $|f(k)|$ consists of a series of peaks, indexed by $\ell$, that reach their respective maximum at $k=\varphi_\ell$ and a width given by $\lambda_\ell$.

\section{Code}
The code we developed for the numerical simulations and figures presented in this work can be found in Ref.~\cite{Code}.

\end{document}